\documentclass[a4paper,11pt,oneside]{article}
\usepackage{a4,theorem} \usepackage{amsmath} \usepackage{latexsym}
\usepackage{amssymb} \usepackage{epsfig}
\usepackage{amsfonts}

\numberwithin{equation}{section}

\def\bexe{\begin{exercise}}\def\eexe{\eex\end{exercise}}
\def\bsol{\begin{solution}}\def\esol{\eex\end{solution}}
\def\bexa{\begin{example}}\def\eexa{\eex\end{example}}
\def\brem{\begin{remark}}\def\erem{\eex\end{remark}}

\newcommand{\R}{{\mathbb R}}
\newcommand{\N}{{\mathbb N}}

\def\CC{{\cal C}}

\def\CA{{\cal A}}  
\def\CG{{\cal G}}\def\CH{{\cal H}}\def\CL{{\cal L}}
\def\CS{{\cal S}}
\def\CB{{\cal B}}

\def\uh{\widehat{u}}\def\vh{\widehat{v}}
\def\noi{\noindent}\def\ds{\displaystyle}
\def\pa{{\partial}}\def\lam{\lambda}
\newcommand{\bi}{\begin{itemize}}\newcommand{\ei}{\end{itemize}}
\newcommand{\bce}{\begin{center}}\newcommand{\ece}{\end{center}}
\newcommand{\reff}[1]{(\ref{#1})}
\newcommand{\ov}[1]{{\overline {#1}}}
\newcommand{\ol}[1]{{\overline {#1}}}

\def\ra{\rightarrow}

\newcommand{\barr}{\begin{array}}\newcommand{\earr}{\end{array}}
\newcommand{\bpm}{\begin{pmatrix}}\newcommand{\epm}{\end{pmatrix}}
\newcommand{\ba}{\begin{array}}\newcommand{\ea}{\end{array}}
\def\dd{\, {\rm d}}

\def\er{{\rm e}}

\def\ddt{\frac{\rm d}{{\rm d}t}}

\def\ddr{\frac{\rm d}{{\rm d}r}}
\def\dds{\frac{\rm d}{{\rm d}s}}

\def\supp{{\rm supp}}

\def\nab{\nabla}\def\eex{\hfill\mbox{$\rfloor$}}
\def\na{\nabla}

\def\al{\alpha}

\def\bd{\begin{displaymath}} \def\ed{\end{displaymath}}
\def\ba{\begin{array}} \def\ea{\end{array}}

\newtheorem{theorem}{Theorem}[section]
\newtheorem{definition}[theorem]{Definition}

\newtheorem{lemma}[theorem]{Lemma}

\newtheorem{exercise}[theorem]{Exercise}
\newtheorem{solution}[theorem]{Solution}
\newtheorem{remark}[theorem]{Remark}
\newtheorem{corollary}[theorem]{Corollary}
\newtheorem{example}[theorem]{Example}

\def\Gh{\widehat{G}}\def\Bh{\widehat{B}}
 \def\qed{\hfill
  $\Box$}\def\wt{\widetilde}

\def\uti{\wt{u}}\def\vti{\wt{v}}\def\wti{\wt{w}}
   \def\ddr{\frac{{\rm d}}{{\rm
      dr}}}\def\dds{\frac{{\rm d}}{{\rm ds}}}

\def\ti{\times} \def\Go{\overline{G}} \def\Vo{\overline{V}}
\def\Bo{\overline{B}} \def\fo{\overline{f}} \def\uo{\overline{u}}
\def\Qti{\widetilde{Q}} \def\CGti{\widetilde{\CG}}
\def\pti{\widetilde{p}} \def\Hh{\widehat{\cal H}} \def\fh{\widehat{f}}
\def\gh{\widehat{g}} \def\vn{v^{(n)}} \def\vdn{\dot{v}^{(n)}}
\def\vdm{\dot{v}^{(m)}}  \def\vnl{v^{({n_l})}}
\def\wn{w^{(n)}}   \def\Pn{P^{(n)}}

\begin{document}
\mbox{}\vspace{0.1cm}\begin{center}\Large Well-posedness of some
  initial-boundary-value
  problems for dynamo-generated poloidal magnetic fields\\[1mm]
  [corrected version]\\[4mm]
  \normalsize
  Ralf Kaiser$^1$ and Hannes Uecker$^2$\\[2mm]
  \small $^1$ Fakult\"at f\"ur Mathematik, Physik und Informatik, Universit\"at
  Bayreuth,
  D-95440 Bayreuth, Germany\\
  $^2$ Institut f\"ur Mathematik, CvO Universit\"at Oldenburg,
  D-26111 Oldenburg, Germany\\
  ralf.kaiser@uni-bayreuth.de, hannes.uecker@uni-oldenburg.de\\
  \normalsize  December, 2012
\end{center}
\begin{abstract}
  Given a bounded domain $G \subset \R^d$, $d\geq 3$, we study smooth
  solutions of a linear parabolic equation with non-constant
  coefficients in $G$, which at the boundary have to $C^1$-match with
  some harmonic function in $\R^d \setminus \ov{G}$ vanishing at
  spatial infinity.

  This problem arises in the framework of magnetohydrodynamics if
  certain dynamo-generated magnetic fields are considered: For
  example, in the case of axisymmetry or for non-radial flow fields,
  the poloidal scalar of the magnetic field solves the above problem.

  We first investigate the Poisson problem in $G$ with the above
  described boundary condition as well as the associated eigenvalue
  problem and prove the existence of smooth solutions. As a by-product
  we obtain the completeness of the well-known poloidal ``free decay
  modes" in $\R^3$ if $G$ is a ball. Smooth solutions of the evolution
  problem are then obtained by Galerkin approximation based on these
  eigenfunctions.

  \vspace{5mm}

  \noi Key words: magnetohydrodynamics, dynamo theory, poloidal field,
  harmonic field.
\end{abstract}
\section{Introduction}
We are concerned in this paper with the following
initial-boundary-value problem:
\begin{subequations}
  \label{1.1}
  \begin{alignat}{2}
    \pa_t u - a\Delta u &= b\cdot \nab u + c\, u &&\qquad \mbox{ in } G \ti \R_+,\\
    \Delta u &= 0 &&\qquad \mbox{ in } \Gh \ti \R_+,\\
    u \mbox{ and } \na u &\mbox{ continuous } &&\qquad \mbox{ in } \R^d \ti \R_+, \\
    u(x,t) &\ra u_\infty(t) &&\qquad \mbox{ for } |x| \ra \infty,\, t \in \R_+ , \\
    u(\cdot,0) &= u_0 &&\qquad \mbox{ on } G \ti \{t = 0\}.
  \end{alignat}
\end{subequations}
Here, $G\subset\R^d$, $d\geq 3$ is a bounded domain with
(sufficiently) smooth boundary $\pa G$, and $\Gh := \R^d \setminus
\Go$. The scalar-valued coefficients $a$ and $c$, and the
vector-valued coefficient $b$ are sufficiently smooth functions of
$x\in G$ and $t \in \R_+$; $a$ is, moreover, bounded from below by
$a_0 > 0$. The asymptotic behaviour of solutions at spatial infinity
is described by the (given) function $u_\infty: \R_+ \ra \R$, and the
initial-value $u_0$ is prescribed on $G$ only.

Problem (\ref{1.1}) arises in the context of magnetohydrodynamic
dynamo theory: the generation of a magnetic field $B$ by motion of a
liquid conductor (of magnetic diffusivity $\eta > 0$) according to some
prescribed flow field $v$ is described by the induction equation
(cf.~\cite{moffatt78})
\begin{gather}
  \pa_t B = \na \ti (v \ti B) - \na \ti (\eta \na \ti B), \quad \na
  \cdot B = 0.
  \label{1.2}
\end{gather}
Equation (\ref{1.2})$_1$ constitutes a system of parabolic equations
for the magnetic field components. In general, the flow field couples
these components in a nontrivial way which makes the question for
``dynamo solutions", i.e.\ solutions which do not decay in time,
difficult to answer. Only in special situations does a field component or a
related scalar quantity decouple, and a general decay result, a
so-called antidynamo theorem, may be obtained. For instance, if the
conductor fills a ball $B_R \subset \R^3$ with radius $R$, if the
conductivity is radially symmetric, and if the flow field has no
radial component the quantity $P:= B\cdot x$ satisfies the scalar
problem (cf.\ appendix A or \cite{kai07}):
\begin{subequations}
  \label{1.3}
  \begin{alignat}{2}
    \pa_t P - \eta \Delta P &= - \na \cdot (v P) &&\qquad \mbox{ in } B_R \ti \R_+,\\
    \Delta P &= 0 &&\qquad \mbox{ in } \Bh_R \ti \R_+,\\
    P \mbox{ and } \na P &\mbox{ continuous } &&\qquad \mbox{ in } \R^3 \ti \R_+, \\
    P(x,t) &=O (|x|^{-2}) &&\qquad \mbox{ for } |x| \ra \infty,\, t \in \R_+ , \\
    P(\cdot,0) &= P_0\, ,\;\; \langle P_0\rangle = 0 &&\qquad \mbox{
      on } B_R \ti \{t = 0\}.
  \end{alignat}
\end{subequations}
Note that conditions (\ref{1.1}b) and (\ref{1.1}d) with $u_\infty
\equiv 0$ imply the spatial decay condition $P(x,t) = O(|x|^{-1})$
(see Appendix C). The stronger condition (\ref{1.3}d) is a consequence
of the additional zero-spherical-mean condition 
$$
\langle P_0 \rangle := \frac{1}{4 \pi r^2} \int_{S_r} P_0 \dd s = 0,\qquad r \in (0,R) ,
$$ 
on the initial value. This condition is preserved by eqs.\ (\ref{1.3}a,b) and
holds, consequently, for $P$ on $\R^3 \ti \R_+$. Equation (\ref{1.3}b)
describes a vacuum field outside the conductor and condition
(\ref{1.3}c) guarantees a continuous magnetic field throughout space.

Another instance is the axisymmetric dynamo problem in ordinary space
$\R^3$. Again, a scalar quantity describing the poloidal part of the
magnetic field decouples and, if reformulated in $\R^5$, is precisely
a solution of problem (\ref{1.1}) (see appendix B). Note that the
conductor is here assumed to be axisymmetric but need not be a
ball. It is this application which motivates the investigation of
problem (\ref{1.1}) in more than 3 dimensions and in domains more
general than balls.

The focus of dynamo theory is less on existence theorems than on decay
results for the magnetic field under certain restrictions on the
magnetic field and/or the flow field, thus excluding dynamo action
under these restrictions. However, proving decay results requires
sometimes the solution of an auxiliary problem. For instance, in
proving a ``non-radial velocity theorem" for solutions of problem
(\ref{1.3}) one needs positive solutions of an auxiliary problem of
type (\ref{1.1}); and it is this application which requires a non-zero
asymptotic condition like (\ref{1.1}d) (cf.~\cite{kai07}). Similarly
in the axisymmetric problem, Backus makes use of solutions of an
auxiliary problem (cf.~\cite{backus57}). He made the existence of such
solutions plausible by physical arguments but could not establish them
rigorously. It is the aim of the present paper to prove rigorously the
existence of smooth solutions of problem (\ref{1.1}).

A problem closely related to (\ref{1.1}) has been treated in 
\cite{sml86}. It is inspired by the dynamo problem with
plane symmetry (which means $d=2$) and differs from (\ref{1.1}) in
that $c\equiv 0$ and by a different asymptotic condition at spatial
infinity. The authors treat this problem and two related ones in
arbitrary dimensions and prove existence of solutions and, moreover,
some decay results. However, this problem does not precisely meet the
requirements of the non-radial problem (\ref{1.3}) nor those of the
axisymmetric problem. Moreover, only weak solutions are established,
whose behaviour at the boundary remains open.

The basic idea of our treatment is to consider (\ref{1.1}) as
parabolic problem in a bounded domain with non-local boundary
condition, and to carry over the well-esta\-blished methods for linear
parabolic equations with standard boundary conditions such as
Dirichlet's or Neumann's boundary condition to our situation (cf.\
e.g.~\cite{evans}). In section 2 we solve weakly a Poisson-type problem
related to (\ref{1.1}) in all space. The regularity of the weak
solution follows with standard arguments in $G$ and $\Gh$, only at
$\pa G$ we need special considerations. In section 3 we treat the
corresponding eigenvalue problem, introduce the associated Fourier
series, and characterize the elements of various function spaces which
are useful in the remainder of the paper by the behaviour of their Fourier
coefficients. In section 4 problem (\ref{1.1}) is solved by a Galerkin
procedure and, for sufficiently smooth data and suitable compatibility conditions,
the smoothness of the obtained solution is established. In two appendices the relation
between problem (\ref{1.1}) and the non-radial-flow as well as the
axisymmetric problem is elucidated. A third appendix collects some
facts about exterior harmonic functions, and clarifies the relation
between different kinds of spatial decay conditions -- a topic about
which there has been some debate in the literature. Finally, in a fourth
appendix the completeness of the so-called poloidal free decay modes
is proved: a commonly believed property which to our knowledge, however,
has never been proved. 
\section{A Poisson problem}
We establish in this section smooth solutions of the following Poisson
problem in all space with suitable right-hand side $f$:
\begin{subequations}
  \label{2.1}
  \begin{alignat}{2}
    - \Delta u &= f &&\qquad \mbox{ in } G ,\\
    \Delta u &= 0 &&\qquad \mbox{ in } \Gh ,\\
    u \mbox{ and } \na u &\mbox{ continuous } &&\qquad \mbox{ in } \R^d , \\
    u(x) &\ra 0 &&\qquad \mbox{ for } |x| \ra \infty .
  \end{alignat}
\end{subequations}
To obtain a weak formulation of problem (\ref{2.1}) let us multiply
(\ref{2.1}a) and (\ref{2.1}b) with a test function $v\in C_0^\infty
(\R^d)$ and integrate over $G$ and $B_R\setminus \Go$, $\Go \subset
B_R$, respectively. Assuming $\pa G \in C^1$ one obtains after
integration by parts:
\begin{subequations}
  \label{2.2}
  \begin{gather}
    \int_G (\nab u\cdot \nab v - f v)\dd x - \int_{\pa G} n \cdot \na u\, v \dd s = 0 , \\
    \int_{B_R\setminus \Go} \nab u \cdot \nab v\dd x + \int_{\pa G} n
    \cdot \na u\, v \dd s - \int_{S_R} \frac{x}{|x|} \cdot \na u\, v
    \dd s = 0.
  \end{gather}
\end{subequations}
Here $S_R$ denotes a sphere with radius $R$ and $n$ is the exterior
unit normal at $\pa G$. Adding up (\ref{2.2}a) and (\ref{2.2}b), one
finds in the limit $R\ra \infty$ that
\begin{equation}
  \label{2.3}
  \int_{\R^d} \na u \cdot \na v \dd x = \int_G f\, v \dd x 
\end{equation}
In view of (\ref{2.3}) it is reasonable to consider functions
satisfying the ``finite energy condition" (cf.\ remark A.1)
\begin{equation}
  \label{2.4}
  \int_{\R^d} |\na v|^2 \dd x =: \| v \|^2_{\CH} < \infty.
\end{equation}
Condition (\ref{2.4}) together with condition (\ref{2.1}d) motivate
the definition of the real Hilbert space
$$
\CH_0 := \mbox{clos} \{C_0^\infty (\R^d), \, \| \cdot \|_\CH \}
$$
with scalar product $(v, w)_\CH := \int_{\R^d} \na v \cdot \na w \dd
x$. Observe that $v \in \CH_0$ is locally square-integrable. In fact,
the Gagliardo-Nirenberg-Sobolev-inequality (cf.~\cite[p263]{evans})
implies
\begin{equation}
  \label{2.5}
  \| v\|_{L^p (\R^d)} \leq C \| v \|_{\CH}\, , \qquad p = \frac{2\, d}{d - 2}
\end{equation}
for any $v \in \CH_0$ and a constant $C$ depending only on
$d$. Combining (\ref{2.5}) with H\"older's inequality yields for any
$K \Subset \R^d$:
\begin{equation}
  \label{2.6}
  \| v\|_{L^2 (K)} \leq |K|^{1/d} \| v\|_{L^p (K)} \leq C |K|^{1/d}  \| v \|_{\CH}\, .
\end{equation}
So, defining (for later use)
$$
\CH := \{ v \in H^1_{loc} (\R^d) : \| v \|_\CH < \infty \} ,
$$
there holds clearly $\CH_0 \subset \CH$. Inequality (\ref{2.6})
implies, in particular,
\begin{equation}
  \label{2.7}
  \| v\|_{L^2 (G)} \leq C_G  \| v \|_{\CH}\, , \qquad v \in \CH_0
\end{equation}
with $C_G := C |G|^{1/d}$; thus, $v \in \CH_0$ yields in eq.\
(\ref{2.3}) with $f \in L^2 (G)$ a finite right-hand side as well.

A function $u \in \CH_0$ is now called a weak solution of problem
(\ref{2.1}) with $f \in L^2(G)$ iff (\ref{2.3}) holds for all $v \in
\CH_0$. Rewriting (\ref{2.3}) in the form
\begin{equation}
  \label{2.11}
  (u, v)_{\CH} = (f, v)_{L^2(G)} \qquad \mbox{ for any } v \in \CH_0
\end{equation}
and noting that $(f, \cdot)_{L^2(G)}$ defines a bounded linear functional on
$\CH_0$ (due to (\ref{2.7})) the existence of a unique weak solution
follows immediately from the Riesz representation theorem.

Concerning regularity of the weak solution let us define $\fh: \R^d
\ra \R$ by $\fh = f$ on $G$ and $\fh = 0$ on $\Gh$. It is a standard
result about interior regularity (cf.\ e.g.\ \cite[p309f]{evans})
that $\fh \in H^k (\R^d)$ implies $u\in H_{loc}^{k+2} (\R^d)$,
$k\in\N_0$. This result means in particular:
\begin{subequations}
  \label{2.12}
  \begin{gather}
    u \in H_{loc}^2 (\R^d), \\
    f\in H^k(G) \Rightarrow u\in H_{loc}^{k+2} (G), \qquad k \in \N_0, \\
    u \in C^\infty (\Gh) .
  \end{gather}
\end{subequations}
With the usual Sobolev embeddings (\ref{2.12}b) implies $u \in C^2
(G)$ if $f\in H^k(G)$, $k>d/2$. So, choosing suitable test functions
in (\ref{2.3}) we find $u$ satisfying (\ref{2.1}b) in $\Gh$ and, for
sufficiently regular $f$, (\ref{2.1}a) in $G$. At $\pa G$, however, we
need some finer considerations.

As usual we flatten $\pa G$ locally by means of a diffeomorphism $\Phi
: U \ra W$, $x \mapsto y$ with $\Phi (x_0) =0$, $x_0 \in U \cap \pa
G$, $\det D \Phi = 1$ (cf.\ e.g.\ \cite[p626f]{evans}). Setting $v(y) :=
u(\Phi^{-1} (y))$ and $g(y):= f(\Phi^{-1} (y))$, $\Phi$ transforms
eqs.\ (\ref{2.1}) localized on $U$ into
\begin{subequations}
  \label{2.13}
  \begin{alignat}{2}
    - L v &= g &&\qquad \mbox{ in } W^- ,\\
    L v &= 0 &&\qquad \mbox{ in } W^+ ,\\
    v \mbox{ and } \na v &\mbox{ continuous } &&\qquad \mbox{ in } W .
  \end{alignat}
\end{subequations}
Here, 
$$
L(\cdot) = \sum_{i,j=1}^d \pa_{y_i} (a_{ij} \pa_{y_j}\, \cdot)
$$
is a uniformly elliptic operator with coefficients determined by
$\Phi$, $W^- := W \cap \{y_d<0\} = \Phi (U\cap G)$, $W^+ := W \cap
\{y_d>0\} = \Phi (U\cap \Gh)$, and $W^0 := W \cap \{y_d=0\} = \Phi
(U\cap \pa G)$. A function $v \in H^1 (W)$ is called a weak solution
of problem (\ref{2.13}) with $g \in L^2(W^-)$ iff
$$
\sum_{i,j=1}^d \int_W a_{ij} \pa_{y_j} v\, \pa_{y_i} w \dd y =
\int_{W^-} g\, w \dd y
$$
for any $w \in H_0^1 (W)$. Defining ``non-isotropic" Sobolev spaces
$H_{tan}^k (W)$, $k\in \N_0$ by
$$
H_{tan}^k (W) := \{ v: W\ra \R\, |\, D^\alpha v \in L^2 (W) \mbox{ for
  any multiindex $\al$ with } |\al| \leq k,\, \al_d = 0 \}
$$
with norm
$$
\|v\|^2_{H_{tan}^k (W)} := \sum_{\underset{\al_d =0}{|\al| \leq k}} \|
D^\al v \|^2_{L^2(W)} ,
$$
the regularity of $v$ over $W^0$ is characterized by
\begin{lemma}
  Let $g \in H_{tan}^k (W^-)$ with $k >(d+1)/2$, let $v$ be a weak
  solution of {\rm (\ref{2.13})}, and let $r$ be so small that the
  cylinder $V := \{ y =(y',y_d) \in \R^d : |y'|< r,\, |y_d|<2r \}
  \Subset W$. Then $v \in C^1 (V_{1/2})$ with $V_{1/2} := V\cap
  \{|y_d| < r\}$ and we have the estimate
  \begin{equation}
    \label{2.14}
    \|v\|_{C^1(\Vo_{1/2})} \leq C\big(\|g\|_{H_{tan}^k (W^-)} + \|v\|_{L^2(W)}\big) 
  \end{equation}
  with $C$ depending on $W$, $V$, $k$, and $d$.
\end{lemma} {\textsc{Proof:}} We prove first the estimate
\begin{equation}
  \label{2.15}
  \|v\|^2_{C^1(\Vo_{1/2})} \leq C \sum_{i=0,1,2}\; \sum_{
    \underset{\al_d =0}{|\al|\leq k}} \|\pa_{y_d}^{\,i} D^\al v \|_{L^2(V)}^2 
\end{equation}
for functions $v \in C^{k+2} (\Vo)$ by applying Sobolev embeddings
separately for $y'$ and $y_d$. Let us start with a 1-dimensional
estimate on the interval $(-r,r)$. For functions $f \in C^1((-2r,2r))$
we have
\begin{gather*}
\begin{split}
|f(y)| &\leq \frac{1}{2r} \int_{y-r}^{y+r} |f(y)-f(x)| \dd x +
\frac{1}{2r} \int_{y-r}^{y+r} |f(x)| \dd x \\
&\leq \int_{-2r}^{2r}
|f'(x)| \dd x +\frac{1}{2r} \int_{-2r}^{2r} |f(x)| \dd x .
\end{split}
\end{gather*}
This implies that
\begin{equation}
  \label{2.16}
  \max_{|y| \leq r} |f|^2 \leq C \Big(\int_{-2r}^{2r} |f'(x)|^2 \dd x + \int_{-2r}^{2r} |f(x)|^2 \dd x \Big)
\end{equation}
for some constant $C = C(r)$. On the other hand, Sobolev embeddings
imply for $k > (d+1)/2$
\begin{equation}
  \label{2.17}
  \|f\|^2_{C^1(\Bo'_r)} \leq C\, \|f\|^2_{H^k(B'_r)} = C \sum_{|\al|\leq k} \int_{B'_r} |D^\al f (y')|^2 \dd y' 
\end{equation}
for some constant $C = C(k,d,r)$ and $B'_r := \{y' \in \R^{d-1} : |y'|
<r \}$. Combining (\ref{2.16}) and (\ref{2.17}) we obtain
\begin{gather}
  \begin{split}
    \max_{|y_d| \leq r} \|v(\cdot ,y_d)\|^2_{C^1(\Bo'_r)} &\leq C
    \sum_{i=0,1}\; \sum_{\underset{\al_d = 0}{|\al|\leq k}}
    \int_{-2r}^{2r} \int_{B'_r} |\pa_{y_d}^{\,i} D^\al v(y', y_d) |^2 \dd y' \dd y_d \\
    &= C \sum_{i=0,1}\; \sum_{ \underset{\al_d =0}{|\al|\leq k}}
    \|\pa_{y_d}^{\,i} D^\al v \|_{L^2(V)}^2 ,
  \end{split}
\end{gather}
and analogously
$$
\max_{|y_d| \leq r} \|\pa_{y_d} v(\cdot ,y_d)\|^2_{C^0(\Bo'_r)} = C
\sum_{i=1,2}\; \sum_{ \underset{\al_d =0}{|\al|\leq k}}
\|\pa_{y_d}^{\,i} D^\al v \|_{L^2(V)}^2 .
$$
This is (\ref{2.15}). Now let $v \in L^2 (V)$ with bounded right-hand
side in (\ref{2.15}). It follows with standard arguments that $v$ has
a representative in $C^1(\Vo_{1/2})$.

Next, let $\gh$ be the trivial extension of $g$ on $W$. $g \in
H^k_{tan} (W^-)$ implies then $\gh \in H^k_{tan} (W)$, i.e.\ $D^\al
\gh \in L^2 (W)$ for $|\al| \leq k$, $\al_d = 0$. From interior
regularity for weak solutions of (\ref{2.13}) follows $D^\beta v \in
L^2 (V)$ for $|\beta| \leq k+2$, $\beta_d \leq 2$ and the estimate
\begin{gather}
  \begin{split}
    \sum_{\underset{\beta_d \leq 2}{|\beta|\leq k+2}} \int_V |D^\beta v |^2 \dd y &\leq C\bigg( \sum_{\underset{\al_d =0}{|\al|\leq k}} \int_W | D^\al \gh |^2 \dd y + \|v\|_{L^2 (W)}^2 \bigg) \\
    &= C \Big( \|g\|_{H^k_{tan} (W^-)}^2 + \|v\|_{L^2(W)}^2 \Big)
  \end{split}
\end{gather}
with $C$ depending on $W$, $V$, $k$, and $d$. Together with
(\ref{2.15}) this proves (\ref{2.14}).  \qed
\begin{remark} {\rm Observe for later use that lemma 2.1 applies
    in particular to the case $g:= \lambda\, v|_{W^-}$ with $\lambda \in
    \R$. In fact, given any $k\in \N$ we find $v|_{W^-} \in
    H^{2k}_{tan} (\widetilde{W}^-)$ for some $V \Subset \widetilde{W}
    \Subset W$ just by iterating the interior regularity argument. So,
    lemma 2.1 applies to this eigenvalue problem as well and yields
    $C^1$-smoothness of solutions over the boundary. }
\end{remark}
Concerning the original problem (\ref{2.1}), lemma 2.1 implies for
$\pa G \in C^{k+2}$ (i.e.\ $\Phi \in C^{k+2}$) and $f\in H^k (G)$, $k>
(d+1)/2$ that $u \in C^1$ over $\pa G$.
Collecting the foregoing results we have
\begin{theorem}[Solution of the Poisson problem]
  Let $G \subset \R^d$, $d \geq 3$ be a bounded domain with
  boundary $\pa G$ and $f\in L^2 (G)$.  The Poisson problem {\rm
    (\ref{2.1})} has then a unique weak (in the sense of eq.\ {\rm
    (\ref{2.3}))} solution $u \in H^1_{loc} (\R^d)$. Moreover, if $\pa
  G \in C^{k+2}$ and $f\in H^k(G)$ with $k>(d+1)/2$, then $u$ is a
  classical solution, i.e.\ $u \in C^1 (\R^d) \cap C^2(G \cup \Gh)$,
  satisfying pointwise eqs.\ {\rm (\ref{2.1})}.
\end{theorem}
\begin{remark} {\rm The condition $d\ge 3$ is crucial for theorem 2.3. 
In lower dimensions problem (\ref{2.1}) is overdetermined; to obtain nontrivial solutions 
one has to relax condition (\ref{2.1}c) or (\ref{2.1}d).
As an example in $\R$ consider the problem $-u''=1$ in $G :=(-1,1)$, 
$u''=0$ in $\Gh$, and $\lim_{|x|\ra \infty}u(x)=0$. The only 
continuous solution is $u(x)=\frac 1 2 -\frac 1 2 x^2$ in $G$ and 
$u\equiv 0$ in $\Gh$, which is not $C^1$ over $\pa G$. Similarly, 
in $\R^2$, the only continuous solution of 
$-\Delta u=1$ in $G := \{x \in \R^2: |x| < 1\}$, $\Delta u=0$ in $\Gh$, 
$\lim_{|x|\ra \infty}u(x)=0$ is $u(x)=\frac 1 4-\frac 1 4 |x|^2$ 
in $G$ and  $u\equiv 0$ in $\Gh$. 

On the other side, $C^1$-solutions of these problems exhibit linear (in $\R$)
or logarithmic (in $\R^2$) growth at infinity. $d=3$ is the lowest dimension, 
in which $C^1$-solutions vanish at infinity. In fact, the analogous problem in $\R^3$
has the $C^1$-solution $u(x)=\frac 1 2 - \frac 1 6 |x|^2$ in $G$ and $u(x)=\frac 1 3 |x|^{-1}$ 
in $\Gh$. }
\end{remark}
\begin{remark}{\rm 
Concerning the asymptotic behaviour of $u$ for large $x$ we refer to 
well-known facts about exterior harmonic functions (cf.\ appendix C). 
In particular, condition (\ref{2.1}d) implies the representation (\ref{C.3}), 
which yields the asymptotics  
\begin{equation}
\label{2.18}
u = Y_0\, |x|^{2-d} + O(|x|^{1-d}), \quad Y_0 \in \R \qquad \mbox{ for }
 \, |x| \ra \infty .
\end{equation}
The decay is faster (viz.\ $O(|x|^{1-d})$), if $f$ has vanishing 
mean over $G$. In fact, one obtains from (\ref{2.1}), (\ref{2.18}), 
and Gauss's theorem:
\begin{equation}
\label{2.19}
|G|\, \fo = - \int_G \Delta u \dd x = - \int_{\pa G} n \cdot \na u \dd s 
= - \int_{S_R} \frac{x}{|x|}\cdot \na u \dd s = (d-2) |S_1| Y_0 .
\end{equation}
Here, $\fo := \frac{1}{|G|} \int_G f \dd x$ and $S_R \subset \Gh$; the 
last equality arises in the limit $R\ra \infty$. So, $\fo=0$ implies $Y_0=0$. 

Let us note as an aside that in general $\uo = 0$ can only be achieved
at the expense of relaxing (\ref{2.1}d) to $u(x) = O(1)$ for $|x|\ra
\infty$. In this case the function $u - \uo$ with $u$ being a solution
of (\ref{2.1}) is obviously a zero-mean solution ( see also remark
2.6).  }
\end{remark}
\begin{remark}{\rm
If $G$ is a ball $B_R$ it makes sense to consider spherically symmetric solutions of problem (\ref{2.1}), i.e.\ $\langle u\rangle (r) = 0$ for any $r>0$. Obviously, $\langle u \rangle = 0$ implies $\langle f \rangle = 0$; on the other hand, any solution of (\ref{2.1}) with spherically symmetric $f$ is spherically symmetric. This follows from the unique solvability of the sub-problem for the spherical mean arising from (\ref{2.1}):\footnote{Prime means differentiation with respect to $r$.}
\begin{gather*}
- \big(\langle u\rangle'' + (d-1)/r\, \langle u\rangle' \big) = \langle f \rangle , \qquad 0\leq r \leq R\, , \\
\langle u \rangle (R) = Y_0/R^{d-2} , \qquad \langle u \rangle' (R) = (2-d) Y_0/R^{d-1} .
\end{gather*}
Observe here the representation (\ref{C.3}), which implies $\langle u \rangle  = Y_0\, r^{2-d}$ for $r> R$, and (\ref{2.19}), which implies $Y_0 =0$ if $\langle f \rangle = 0$.

Note, finally, that spherically symmetric solutions decay at least
like $|x|^{1-d}$ for large~$x$.  }
\end{remark}
\section{The Eigenvalue problem}
We treat in this section the eigenvalue problem corresponding to the
Poisson problem (\ref{2.1}) of the previous section:
\begin{subequations}
  \label{3.1}
  \begin{alignat}{2}
    - \Delta u &= \lambda u &&\qquad \mbox{ in } G ,\\
    \Delta u &= 0 &&\qquad \mbox{ in } \Gh ,\\
    u \mbox{ and } \na u &\mbox{ continuous } &&\qquad \mbox{ in } \R^d , \\
    u(x) &\ra 0 &&\qquad \mbox{ for } |x| \ra \infty .
  \end{alignat}
\end{subequations}
According to theorem 2.3 we have for any $f\in L^2(G)$ a unique weak
solution $u \in \CH_0$ of (\ref{2.1}), defining thus a Green
operator
$$
\CGti : L^2(G) \ra \CH_0\, , \quad f\mapsto u\, .
$$ 
Using (\ref{2.7}), (\ref{2.11}) we obtain \footnote{The symbols $L^2$
and $H^k$ without specified domain mean always $L^2 (G)$ and $H^k (G)$, respectively.}
\begin{equation}
\label{3.1A}
\|\wt{\CG} f\|_\CH^2 = (u,u)_\CH = (f,u)_{L^2} \le
\|f\|_{L^2}\|\wt{\CG} f\|_{L^2} \le C_G \|\wt{\CG} f\|_{\CH} \|
f\|_{L^2}
\end{equation}
and
$$
(f,\wt{\CG} f)_{L^2} = (f,u)_{L^2} = (u,u)_\CH = \|\wt{\CG} f\|_\CH^2
\geq 0.
$$
Therefore, $\CGti$ is a bounded linear operator between Hilbert
spaces, which is, furthermore, positive and hence
symmetric. Restricting $u$ on $G$ one obtains the operator
$$
\CG : L^2 (G) \ra L^2 (G)\, ,\quad f \mapsto u|_G\, ,
$$
which is likewise bounded and symmetric, and, moreover, compact due to
the Rellich-Kondrachov theorem and the observation $\{ u|_G : u \in
\CH_0\} = \{u|_G : u \in \CH \} = H^1 (G)$. The spectral theorem for
symmetric compact operators in Hilbert spaces establishes now a
complete (in $L^2 (G)$) orthonormal system $\{ v_n : n \in \N\}$ of
eigenvectors of $\CG$:
\begin{equation}
  \label{3.2}
  \CG\, v_n = \lambda^{-1}_n v_n\, , \qquad n \in \N\, ,
\end{equation}  
with real, positive eigenvalues $\lambda^{-1}_n$ of finite
multiplicity and $\lim_{n \ra \infty} \lambda^{-1}_n = 0$. In order to
solve the original problem let us define the ``harmonic extension"
$\vti_n := \lambda_n \CGti v_n$ of $v_n$ on $\R^d$.\footnote{In this
  section a quantity with tilde always means the ``harmonic extension"
  on $\R^d$ of a quantity defined on $G$.} By definition the $\vti_n$
are weak solutions of the Poisson problem (\ref{2.1}) with $f:= v_n$
and $u = \lambda^{-1} \vti_n$; thus, eq.\ (\ref{2.11}) takes now the
form
\begin{equation}
  \label{3.3}
  (\vti_n , v)_\CH = \lambda_n (v_n, v)_{L^2}  \qquad \mbox{ for any } v \in \CH_0\, ,
\end{equation}  
i.e.\ the pair $(\vti_n , \lambda_n)$ is a weak solution of the
eigenvalue problem (\ref{3.1}). 
Note, finally, the uniqueness of the harmonic extensions in $\CH_0$,
which is implied by the uniqueness of weak solutions.

Concerning regularity we have $\vti_n \in C^\infty (G\cup \Gh) \cap
C^1 (\R^d)$, which follows from (\ref{2.12}c), iterating
(\ref{2.12}b), and remark 2.2. So, $(\vti_n, \lambda_n)$ is also a
classical solution of problem (\ref{3.1}).

\noi We summarize these results in
\begin{theorem}[Solution of the eigenvalue problem]
  Let $G \subset \R^d $, $d\geq 3$ be a bounded domain with
  $C^k$-boundary, $k > (d+5)/2$. The eigenvalue problem {\rm
    (\ref{3.1})} then has a countable set of eigensolutions
  $\{(\vti_n, \lambda_n) : n \in \N \}$ satisfying {\rm (\ref{3.3})},
  and their restrictions $\{v_n : n\in \N\}$ constitute an orthonormal
  basis of $L^2 (G)$.
\end{theorem}

Powers of the inverse Green operator and their domains of definition
turn out to provide the right setting for the solution of the
evolution problem in Section 4. The elements of these spaces can be
characterized by the decay behaviour of their Fourier coefficients
when expanded in the above eigenfunctions. This motivates the
\begin{definition}
  Let $\{v_n : n \in \N\}$ be the complete orthonormal system with
  associated eigenvalues $\lambda_n$ according to theorem {\rm 3.1},
  and $\al \in \R$. We define then the space of ``formal series"
$$
\CS := \bigg\{\sum_{n=1}^\infty c_n v_n :\, c_n \in \R \bigg\} ,
$$
with non-negative functional
$$
\|\cdot\|_\al :\, \CS \ra [0,\infty]\, , \quad \sum_{n=1}^\infty c_n
v_n \mapsto \bigg(\sum_{n=1}^\infty \lambda_n^{2\al} |c_n|^2
\bigg)^{1/2},
$$ 
linear mapping
$$
\CA^\al : \CS \ra \CS\, , \quad \sum_{n=1}^\infty c_n v_n \mapsto
\sum_{n=1}^\infty \lambda_n^\al c_n v_n ,
$$
and subspaces
$$
D_\al :=D(\CA^\al) = \{ v \in \CS : \|v\|_\al < \infty \} \subset \CS
.
$$
\end{definition}
Obviously, $\CA^\al $ maps $D_\al$ into $L^2 (G)$. Furthermore, there
is $D_\al \subset D_\beta$ if $\al \geq \beta$ and $D_0 =
L^2(G)$. Thus, if $\al \geq 0$, there is $D_\al \subset L^2(G)$ and $v
\in D_\al$ has the representation
\begin{equation}
  \label{3.5}
  v = \sum_{n=1}^\infty (v_n, v)_{L^2}\, v_n\, .
\end{equation}
With the pairing
\begin{equation}
\label{3.5a}
\langle w,v\rangle := \sum_{n=1}^\infty d_n c_n \quad \mbox{for }
w = \sum_{n=1}^\infty d_n v_n \in D_{-\al} ,\; v =
\sum_{n=1}^\infty c_n v_n \in D_\al,
\end{equation}
$D_{-\al}$ is the dual space of
$D_\al$.

Applying $\CA$ on (finite) linear combinations of $v_n$ we find $\CA =
\CG^{-1}$, and $D(\CA)$ turns out to be the maximal domain of
definition of $\CG^{-1}$. Similarly, $D_{1/2}$ is related to
$\CH_0$. More precisely, we have
\begin{theorem}
  Let $G \subset \R^d$, $d \geq 3$ be a bounded domain with $C^\infty$-boundary $\pa G$ and
  let $\{ v_n : n\in \N\}$ be the complete orthonormal system defined
  by the eigenvalue problem {\rm (\ref{3.1})}. Let, furthermore, $\CA$ and $D_\al$ be as
  defined in definition {\rm 3.2}, and $\CG$ be the Green operator
  associated to the Poisson problem {\rm (\ref{2.1})}. Then,
  $$
  D_0 = H^0 (G) = L^2 (G) ,
  $$ 
  \begin{equation}
    \label{1}
    D_{1/2} = \{ v|_G : v \in\CH_0 \mbox{ and } v|_{\Gh} \mbox{ is harmonic}\} = H^1 (G) ,
  \end{equation}
  i.e., in particular, any $v \in D_{1/2}$ has a unique harmonic extension $\vti \in
  \CH_0$, and
  \begin{equation}
  \label{2}
      D_1 = \CG(L^2(G)) = \{ v\in H^2 (G): \vti \in H^2_{loc} (\R^d)\} .
  \end{equation}
  Higher order spaces are characterized by
  \begin{equation}
    \label{3}
    D_{k/2} = \Big\{ v \in  H^k (G) : \, \widetilde{\Delta^{i-1} v} \in H_{loc}^2 (\R^d) \mbox{ for }\, i =1,\ldots , [k/2] \Big\}
    \quad k \in \N \setminus \{1\} \, , 
  \end{equation}
  where $\wti \in \CH_0$ again denotes the harmonic extension of a function $w \in D_{1/2} = H^1 (G)$ and $[r] := 
  \max \{j\in \N : j \leq r\}$ is the integer part of $r$. 

On $D_{k/2}$ we have the equivalence of norms:
  \begin{equation}
    \label{4}
    \| \cdot \|_{k/2} \sim \| \cdot\|_{H^k}\, , \qquad k\in \N_0\, .
  \end{equation}
\end{theorem}
To prove the theorem the following lemma is helpful. It improves the regularity result (\ref{2.12}b) and provides the pertinent estimate. 
\begin{lemma} 
Let $G \subset \R^d$, $d \geq 3$ be a bounded domain with $C^{k+2}$-boundary $\pa G$ and $f \in H^k(G)$, $k \in \N_0$. Let, furthermore, $u \in \CH_0$ be the weak solution of problem {\rm (\ref{2.1})}. Then $u \in H^{k+2}(G)$ and we have the bound
\begin{equation}
\label{A.1}
\|u\|_{H^{k+2} (G)} \leq C \|f\|_{H^k(G)} = C \|\Delta u\|_{H^k(G)}
\end{equation}
with a constant $C$ depending on $G$, $k$, and $d$.
\end{lemma}
{\textsc{Proof:}}  The case $k=0$ is already implied by the interior regularity result (\ref{2.12}a). In fact, $u \in H^2_{loc} (\R^d)$ means (see \cite{evans}, p309)
\begin{equation}
\label{A.2}
\|u\|_{H^2(G)} \leq \widehat{C} \Big(\|\fh\|_{L^2(K)} + \|u\|_{L^2(K)}\Big) ,
\end{equation}
where $\widehat{f}$ denotes the trivial extension of $f$ onto $\R^d$ and $K$ some bounded domain such that $G \Subset K$. Combining (\ref{2.6}) with (\ref{3.1A}) we obtain
\begin{equation}
\label{A.3}
\|u\|_{L^2(K)} \leq C_K \|u\|_{\CH} \leq C_K C_G  \|f\|_{L^2(G)}\, ,
\end{equation}
and thus (\ref{A.2}) takes the form 
\begin{equation}
\label{A.3a}
\|u\|_{H^2(G)} \leq C \|f\|_{L^2(G)}\, .
\end{equation}
No boundary regularity is required for this result.

The case $k>0$ needs separate considerations of tangential and normal derivatives at $\pa G$. We refer in the following to the situation, where $\pa G$ has already been flattened as explained in the paragraph before lemma 2.1 and we use the notation introduced there.
So, given $g \in L^2(W^-)$ we assume $v \in H^1(W)$ to be a (weak) solution of
\begin{equation}
\label{A.4}
\sum_{i,j=1}^d \int_W a_{ij} \pa_{y_j} v\, \pa_{y_i} w \dd y =
\int_{W^-} g\, w \dd y
\end{equation}
for any $w \in H_0^1 (W)$. Let $\gh$ be again the trivial extension of $g$ onto $W$. Now we assume higher tangential regularity of $g$, i.e.\ $D^\alpha g \in L^2(W^-)$ for $|\alpha| \leq k$, $\al_d = 0$, which implies $D^\alpha \gh \in L^2(W)$. From interior regularity for weak solutions it follows that $D^\beta v \in L^2(V)$ for $|\beta| \leq k+2$, $\beta_d \leq 2$ and any $V \Subset W$, together with the estimate
\begin{equation}
\label{A.5}
    \sum_{\underset{\beta_d \leq 2}{|\beta|\leq k+2}} \int_{V^-} |D^\beta v |^2 \dd y \leq C\bigg( \sum_{\underset{\al_d =0}{|\al|\leq k}} \int_{W^-} | D^\al g |^2 \dd y + \|v\|_{L^2 (W)}^2 \bigg) .
\end{equation}
As to normal derivatives, note that higher interior regularity implies
\begin{equation}
\label{A.6}
-D^\al \sum_{i,j=1}^d \pa_{y_i} (a_{ij} \pa_{y_j} v) = D^\al g
\end{equation}
to hold a.e. in $W^-$. Writing (\ref{A.6}) with $\al = (0,\ldots, 0,1)$ in the form 
\begin{equation}
\label{A.7}
a_{d d}\, \pa^3_{y_d} v = -\sum_{\underset{i+j < 2d}{i,j = 1}}^d \pa_{y_d} \pa_{y_i} (a_{ij}\, \pa_{y_j} v) - 2\, \pa_{y_d} a_{d d}\,
 \pa^2_{y_d} v - \pa_{y_d}^2 a_{d d}\, \pa_{y_d} v - \pa_{y_d} g\, ,
\end{equation}
we find, by uniform ellipticity, $\pa_{y_d}^3 v$ to be bounded in $W^-$ by the right-hand side of (\ref{A.7}), which is at most of second order in $\pa_{y_d} v$. So, (\ref{A.5}) may be applied and we arrive at
\begin{equation}
\label{A.8}
\int_{V^-} |\pa^3_{y_d} v|^2 \dd y \leq \widetilde{C} \bigg( \sum_{\underset{\al_d =0}{|\al|\leq k}} \int_{W^-} | D^\al g |^2 \dd y + \|v\|_{L^2 (W)}^2 \bigg) .
\end{equation}
The case of arbitrary higher derivatives is now easily proved by induction. So, we find, finally, that (\ref{A.5}) holds without restriction on $\alpha_d$ and $\beta_d$, respectively.

To complete the proof one has, as usual, to cancel the change of variables, to cover $G$ by local patches, to sum up the corresponding local estimates, and to use once more (\ref{A.3}).
\qed

\vspace{1ex}

\noi{\textsc{Proof [of theorem 3.3]:}} The case $k= 0$ is trivial.
To prove (\ref{1}) let $v \in
D_{1/2}$ be decomposed as in (\ref{3.5}), i.e., $v = \sum_{n=1}^\infty
(v_n, v)_{L^2}\, v_n$ with $ \sum_{n=1}^\infty \lambda_n |(v_n,
v)_{L^2} |^2 < \infty$. We define
$$
\vti := \sum_{n=1}^\infty (v_n, v)_{L^2}\, \vti_n
$$
with $\vti_n \in \CH_0$ being the unique harmonic extension of
$v_n$. Computing
\begin{equation}
  \label{3.9}
  \|\vti \|_\CH^2 = \sum_{n,m=1}^\infty (v_m,v)_{L^2} (v,v_n)_{L^2} (\vti_m, \vti_n)_\CH = \sum_{n=1}^\infty \lambda_n |(v_n, v)_{L^2}|^2 , 
\end{equation}
where we used (\ref{3.3}), we find $\vti \in \CH_0$. In order to prove
that $\vti|_{\Gh}$ is harmonic it suffices to show $(\vti, \vh)_\CH =
0$ for any $\vh \in C_0^\infty (G)$ with supp$\,\vh \subset \Gh$; this
follows immediately with (\ref{3.3}):
$$
(\vti,\vh)_{\CH} = \sum_{n=1}^\infty (v_n,v)_{L^2} (\vti_n,\vh)_{\CH}
= \sum_{n=1}^\infty \lam_n (v_n,v)_{L^2} (v_n,\vh)_{L^2} = 0 ,
$$
since $(v_n,\vh)_{L^2} = \int_G v_n\, \vh \dd x = 0$. Therefore, $v
\in \{ v|_G : v \in\CH_0 \mbox{ and } v|_{\Gh} \mbox{ is
  harmonic}\}$. The opposite inclusion follows again with (\ref{3.9}).

The inclusion $\{ v|_G : v \in\CH_0 \mbox{ and } v|_{\Gh} \mbox{ is
  harmonic}\} \subset H^1 (G)$ is obvious; the opposite inclusion
follows with theorem C.1: Let $v \in H^1(G)$, $v_0$ an $H^1$-extension
of $v$ on $\R^d$ with bounded support, $w:= {v_0}|_{\Gh} \in \Hh$ and
$\uh$ the exterior harmonic solution of (\ref{C.2}). The function $u$
defined by $v$ in $G$ and $\uh$ in $\Gh$ is then the sought-after
function $\in \CH_0$. This proves (\ref{1}).

To estimate the $1/2$-norm we supply the above construction with
bounds: Let $\supp\, v_0 \subset K$, then
$$
\|v_0\|_{H^1(\R^d)} \leq C(G,K) \|v\|_{H^1 (G)}\, .
$$
Thus, using the minimizing property of solutions of (\ref{C.2}),
$$
\|\uh \|_{\Hh} \leq \|w\|_{\Hh} \leq \|v_0\|_{\CH} \leq C(G,K)
\|v\|_{H^1 (G)}\, ,
$$
and with (\ref{3.9}):
$$
\|v\|^2_{1/2} = \|\vti\|^2_\CH = \|\na v\|^2_{L^2 (G)} +
\|\uh\|^2_{\Hh} \leq C \|v\|^2_{H^1(G)}\, ,
$$
which is one half of (\ref{4}) for $k=1$. The other half follows
by (\ref{2.7}).

We show next that $D_1 = \CG(L^2)$. Let $u \in \CG (L^2)$ and $f =
\sum_{n=1}^\infty (v_n, f)_{L^2}\, v_n \in L^2$ such that $u = \CG
f$. Computing the coefficients of $u$ we find with (\ref{3.2}):
$$
(v_m, u)_{L^2} = (v_m , \CG f)_{L^2} = \sum_{n=1}^\infty (v_n,
f)_{L^2} (v_m, \CG v_n)_{L^2} = \lambda_m^{-1} (v_m, f)_{L^2}\, .
$$
Thus,
$$
\sum_{m=1}^\infty \lambda_m^2 |(v_m,u)_{L^2}|^2 = \sum_{m=1}^\infty
|(v_m,f)_{L^2}|^2 < \infty\, ,
$$
hence $u \in D_1$. If, on the other hand, 
$$
u = \sum_{n=1}^\infty
(v_n, u)_{L^2}\, v_n \quad \mbox{ and } \quad \sum_{n=1}^\infty \lambda_n^2
|(v_n,u)_{L^2}|^2 < \infty ,
$$
then 
$$
f := \sum_{n=1}^\infty \lambda_n (v_n, u)_{L^2}\, v_n \in L^2
$$ 
is well-defined and we find $\CG f = u$.

The inclusion $\CG(L^2(G)) \subset \{ v \in H^2 (G) : \vti \in H_{loc}^2 (\R^d) \}$ is an immediate consequence
of the $H^2$-regularity of weak solutions. To prove the opposite
inclusion let $w \in H^2(G)$ with harmonic extension $\wti \in \CH_0 \cap H_{loc}^2 (\R^d)$. 
Defining $f := -\Delta w \in L^2$ the
Poisson problem (\ref{2.1}) yields a solution $\uti\in \CH_0 \cap H^2_{loc}(\R^d)$. So,
we have pointwise a.e.\ $\Delta (\wti -\uti) = 0$ in $\R^d$ for $\wti - \uti \in \CH_0 \cap H_{loc}^2 (\R^d)$, 
which means $\wti - \uti$ is harmonic in $\R^d$ (by Weyl's lemma)
and, moreover, $\wti - \uti = 0$ (by Liouville's theorem). Thus, $\wti = \uti$ and, in particular, $w = u = \CG (f)$.

To estimate the $1$-norm of $v \in D(\CA)$ observe that for its
harmonic extension holds: $\vti \in \CH_0 \cap H^2_{loc} (\R^d)$, 
and for the eigenfunctions $v_n$: $v_n \in C^1 (\R^d)$. So, by (\ref{3.3}) 
we can calculate
\begin{equation}
  \label{5}
  -(\lambda_n v_n ,v)_{L^2(G)} = -\int_{\R^d} \na \vti_n \cdot \na \vti \dd x = \int_{\R^d} \vti_n \, \Delta \vti \dd x  = 
  (v_n ,\Delta v)_{L^2(G)}
\end{equation}
and therefore obtain
\begin{equation}
\label{6}
\|v\|_1^2 = \| \CA v\|^2_{L^2} = \sum_{n=1}^\infty \lambda_n^2 
|( v_n,v)_{L^2}|^2 =
\sum_{n=1}^\infty |(v_n,\Delta v)_{L^2}|^2 =\|\Delta v\|_{L^2}^2\, , 
\end{equation}
which implies $\|v\|_1 \leq C \|v\|_{H^2 (G)}$ with a constant $C$ depending only on $d$.
To prove the opposite inequality we combine (\ref{6}) with (\ref{A.1})$_{k=0}$:
$$
\|v\|_1 = \|\Delta v\| \geq \frac{1}{C}\|v\|_{H^2(G)}.
$$
This proves (\ref{4})$_{k=2}$.

The case $k >2$ is proved by induction. Let $v \in D_{k/2 +1}$, $k\in \N$, i.e.\ $\CA v \in D_{k/2}$. By assumption we have $\CA v\in H^k (G)$ and 
\begin{equation}
\label{9}
\widetilde{\Delta^{i-1} \CA v} \in H_{loc}^2 (\R^d) \quad \mbox{for }\, i = 1,\ldots, [k/2]\, .
\end{equation}
(Note that for $v \in D_{3/2}$ condition (\ref{9}) does not yet make sense and can be omitted.)  By (\ref{6}) the condition $\CA v \in H^k(G)$ means $\Delta v \in H^k(G)$, and lemma 3.4 implies $v \in H^{k+2} (G)$. Moreover, we have $\vti \in H^2_{loc} (\R^d)$, which complements condition (\ref{9}). So, we conclude
\begin{equation}
\label{10}
    v \in  \Big\{ v \in  H^{k+2} (G) : \, \widetilde{\Delta^{i-1} v} \in H_{loc}^2 (\R^d) \mbox{ for }\, i =1,\ldots , [k/2]+1 \Big\} .
\end{equation}
To prove the opposite inclusion let $v$ be as in (\ref{10}). We set $w:= \Delta v$ and have by assumption
$$
   w \in \Big\{ v \in  H^k (G) : \, \widetilde{\Delta^{i-1} v} \in H_{loc}^2 (\R^d) \mbox{ for }\, i =1,\ldots , [k/2] \Big\} = D_{k/2}\, .
$$
Computing the $k/2 + 1$-norm of $v$ we find with (\ref{5})
\begin{equation}
\label{11}
\|v\|_{k/2 + 1}^2 = \sum_{n=1}^\infty \lambda_n^k |\lambda_n(v_n,v)_{L^2}|^2 =
\sum_{n=1}^\infty \lambda_n^k|(v_n,w)_{L^2}|^2 =\|w\|_{k/2}^2 < \infty\, ,
\end{equation}
and thus, $v \in D_{k/2 + 1}$. This completes the proof of (\ref{3}).

As to the equivalence (\ref{4}) we proceed likewise by induction. Assuming $v \in D_{k/2 + 1}$, $k \in \N$ we find by (\ref{11}) and by assumption
$$
   \|v\|_{k/2 + 1} = \| \Delta v\|_{k/2} \leq C \|\Delta v\|_{H^k} \leq \widetilde{C} \|v\|_{H^{k+2}} ,
$$
whereas the opposite inequality follows by (\ref{A.1}): 
$$
   \|v\|_{H^{k+2}} \leq C \| \Delta v\|_{H_k} \leq \widetilde{C} \|\Delta v\|_{k/2} = \widetilde{C} \|v\|_{k/2 + 1} .
$$
This completes the proof.
\qed

\begin{remark}{\rm 
Iterating (\ref{5}) one finds on $D_\al$ for integer values $\al$ the following alternative formulation of the $\al$-norm:
$$
\|v\|_k = \|\Delta^k v\|_{L^2(G)}\, , \quad v\in D_k\, ,\quad k \in \N\, , 
$$
and by (\ref{3.9}) for half-integer values:
$$
\|v\|_{k + 1/2} = \|\na \widetilde{\Delta^k v}\|_{L^2(\R^d)}\, , \quad v\in D_{k + 1/2}\, ,\quad k \in \N_0\, . 
$$
}
\end{remark}
\section{The evolution problem}
We solve in this section the evolution problem (\ref{1.1}) by means of
the spaces $D_\al$ provided in the last section.  According to eq.\
(\ref{1}), $v \in D_{1/2}$ has a harmonic extension $\vti$ on
$\R^d$; so, when working with these spaces it is sufficient to
consider problem (\ref{1.1}) on the simpler domain $G\ti
\R_+$. However, a nontrivial asymptotic function $u_\infty$ does not
fit into this framework.  Therefore, in a first step, $u_\infty$ is
eliminated by the time-dependent shift $u - u_\infty := u_s$. In terms
of $u_s$ problem (\ref{1.1}) reads
\begin{subequations}
  \label{4.1}
  \begin{alignat}{2}
    \pa_t u_s - a \Delta u_s &= b\cdot \nab u_s + c\ u_s + f &&\qquad \mbox{ in } G \ti \R_+,\\
    \Delta u_s &= 0 &&\qquad \mbox{ in } \Gh \ti \R_+,\\
    u_s \mbox{ and }& \na u_s \mbox{ continuous } &&\qquad \mbox{ in } \R^d \ti \R_+, \\
    u_s(x,t) &\ra 0 &&\qquad \mbox{ for } |x| \ra \infty,\, t \in \R_+ , \\
    u_s(\cdot,0) &= u_{s\, 0} &&\qquad \mbox{ on } G \ti \{t = 0\}
  \end{alignat}
\end{subequations}
with $f := c\, u_\infty - \ddt u_\infty$.

The ``simplified" problem takes then the form
\begin{subequations}
  \label{4.2}
  \begin{align}
    \dot{v}  &= - a\, \CA\, v + \CB\, v + f ,\\
    v(0) &= v_0,
  \end{align}
\end{subequations}
with the operator $\CA$ as defined in Definition 3.2 and the
lower-order operator $\CB$ defined by $\CB\, v := b \cdot \na v + c\,
v$. Here $v$ is a mapping from $[0,T)$, $T>0$ into some function space
over $G$. As explained above a reasonable such space is $D_{k/2} = H^k
(G)$ with at least $k = 1$ (cf.\ theorem 3.3). Moreover, for $T>0$,
when starting with $v_0 \in D_{1/2}$ and taking into account parabolic
smoothing we expect $v \in L^2((0,T),D_1)$ which means in view of
(\ref{4.2}a) $\dot{v} \in L^2((0,T),L^2 (G))$. This motivates the
\begin{definition}
  Let $T>0$ and $v_0 \in D_{1/2}$. A function $ v \in L^2((0,T),
  D_{1})$ with weak time derivative $\dot{v} \in L^2((0,T),L^2 (G)))$
  satisfying {\rm (\ref{4.2}a)} as equality in $L^2((0,T),L^2 (G))$
  and {\rm (\ref{4.2}b)} as equality in $D_{1/2}$ is called weak
  solution of problem {\rm (\ref{4.2})}.
\end{definition}
Condition (\ref{4.2}b) makes sense for weak solutions due to the
following interpolation result:
\begin{lemma}
  Let $G$ be a bounded domain with smooth boundary, $T>0$, and $k \in
  \N_0$. Let, furthermore, $v \in L^2((0,T), H^{k+1} (G))$ and
  $\dot{v} \in L^2((0,T), H^{k-1} (G))$. Then
$$
v \in C([0,T], H^{k} (G))\, ;
$$
moreover, the mapping $t \mapsto \|v(t)\|^2_{L^2(G)}$ is absolutely
continuous with derivative
\begin{equation}
  \label{4.A}
  \frac{1}{2}\, \ddt\, \| v(t)\|^2_{L^2 (G)} = \langle \dot{v} (t), v(t) \rangle
\end{equation}
for a.e.\ $t \in [0,T]$.
\end{lemma}
For a proof we refer to \cite[p287f]{evans}. We note only in the
case $k=0$ that $D_{-1/2}$ is the dual space of $D_{1/2} = H^1 (G)$,
thus $D_{-1/2} \subset H^{-1} (G)$. $\langle \cdot\, , \cdot \rangle$
denotes the dual pairing as defined by (\ref{3.5a}).
\begin{theorem}[Weak solution of the evolution problem]
  Let $T>0$ and $a,b,c \in C(\Go \ti [0,T])$, $a \geq a_0 >0$. Let,
  furthermore, $v_0 \in D_{1/2}$ and $f \in C([0,T], L^2 (G))$. Then
  problem {\rm (\ref{4.2})} has a unique weak solution $v$.
\end{theorem} {\textsc{Proof:}}
We start with the construction of Galerkin approximations using the
complete system $\{v_\nu : \nu \in \N\}$ from theorem 3.1. Let $P_\nu$
be the orthogonal projection in $L^2 (G)$ onto span$\, \{v_\nu \}$,
$P^{(n)} := \bigoplus_{\nu = 1}^n P_\nu$, and let $\vn(t) \in P^{(n)}
L^2 (G)$ be the unique solution of the following finite-dimensional
initial-value problem
\begin{subequations}
  \label{4.B}
  \begin{align}
    \ddt \vn  &= \Pn (- a\, \CA\, \vn + \CB\, \vn + f)\, ,\\
    \vn(0) &= P^{(n)} v_0\, .
  \end{align}
\end{subequations}
Note that $\Pn$ commutes with $\CA$ but not with $a$ or $\CB$. From
standard results about ordinary differential equations follows $\vn
\in C^1 ([0,T], D_1)$ for any $n \in \N$.

Next we derive some a-priori estimates for $\vn$, uniform in $n$,
which allow to extract a weakly convergent subsequence of the sequence
$(\vn)$ of Galerkin approximations. We first show that
\begin{equation}
  \label{4.3}
  \max_{[0,T]} \| \vn \|_{1/2}^2 \leq C = C[v_0, f; T]\, .
\end{equation}
Taking the scalar product of (\ref{4.B}a) with $\CA \vn$ we obtain
\begin{equation}
  \label{4.4}
  \begin{split}
    \frac{1}{2}\, \ddt \,\|\vn\|^2_{1/2} &= \frac{1}{2}\, \ddt (\CA^{1/2} \vn, \CA^{1/2} \vn)_{L^2} = \big(\CA\, \vn , \ddt \vn \big)_{L^2} \\
    &= ( \CA\, \vn, - a\, \CA\, \vn + \CB\, \vn + f )_{L^2} \\
    &\leq - a_0 \| \vn \|^2_1 + ( \CA\, \vn , \CB\, \vn )_{L^2} + (
    \CA\, \vn , f)_{L^2}\, .
  \end{split}
\end{equation}
Observing that $\CB$ is a bounded operator from $C([0,T], D_{1/2})$
into $C([0,T], L^2(G))$ there is a constant $C_1$ such that
$$
(\CA\, \vn, \CB\, \vn )_{L^2} \leq (C_1 a_0)^{1/2} \|\vn \|_1
\|\vn\|_{1/2} \leq \frac{a_0}{2}\, \|\vn\|_1^2 + \frac{C_1}{2}\, \|\vn
\|_{1/2}^2\, .
$$
Setting $\max_{[0,T]} \| f\|_{L^2}^2 =: C_2 a_0$ we thus obtain
\begin{equation}
  \label{4.5}
  \ddt \,\|\vn\|^2_{1/2} \leq C_1 \| \vn \|^2_{1/2} + C_2\, , 
\end{equation}
and Gronwall's inequality yields
$$
\|\vn\|^2_{1/2} \leq \er^{C_1 t} \|\vn (0)\|^2_{1/2} + \frac{C_2}{C_1}
\big( \er^{C_1 t} - 1\big) \leq \er^{C_1 T} \|\vn_0\|^2_{1/2} +
\frac{C_2}{C_1} \big( \er^{C_1 T} - 1\big)
$$
on $[0,T]$, and hence (\ref{4.3}).

To obtain a bound on $\vn$ in $L^2((0,T), D_1)$ we estimate similarly
to (\ref{4.4}):
\begin{equation}
  \label{4.10}
  \frac{1}{2}\, \ddt \,\|\vn\|^2_{1/2} \leq - a_0 \| \vn \|^2_1 + \frac{a_0}{4} \| \vn \|_1^2 + C_1 \| \vn \|_{1/2}^2 + \frac{a_0}{4} \| \vn \|_1^2 + C_2\, .
\end{equation}
Using (\ref{4.3}) we rewrite (\ref{4.10}) in the form
$$
a_0 \|\vn \|_1^2 \leq -\ddt\, \|\vn \|_{1/2}^2 + 2 (C_1 C + C_2)\, ;
$$
thus, integrating over $[0,T]$ and observing (\ref{4.3}) once more
yields the bound
\begin{equation}
  \label{4.AA}
  \int_0^T \|\vn\|_1^2 \dd t \leq \widehat{C}\, .
\end{equation}
With (\ref{4.AA}) the right-hand side in (\ref{4.B}a) is obviously
bounded in $L^2((0,T),L^2 (G))$, i.e.\ there is $\check{C}$ such that
\begin{equation}
  \label{4.7}
  \int_0^T \big\| \ddt \vn \big\|_{L^2}^2 \dd t \leq \check{C}
\end{equation}
for the sequence of (classical) derivatives $\big( \ddt \vn \big)$.

The bounds (\ref{4.AA}) and (\ref{4.7}) imply that there is a
subsequence $(n_l)$ and functions $v \in L^2((0,T), D_{1})$, $\dot{v}
\in L^2((0,T), L^2 (G))$ such that $v$ is the weak limit of $(\vnl)$
in $L^2((0,T), D_{1})$ and $\dot{v}$ that of $\big(\ddt \vnl\big)$ in
$L^2((0,T), L^2 (G))$, respectively, and moreover $\dot{v}$ is the
(weak) derivative of $v$.

Testing (\ref{4.B}a) with functions $w$ of the form $w(t) = \sum_{\nu=
  1}^m d_\nu (t)\, v_\nu \in C^1 ([0,T], L^2 (G))$, where $m\leq n$
and $d_\nu : [0,T] \ra \R$ are smooth functions, and integrating over
$[0,T]$ yields
\begin{equation}
  \label{4.8}
  \begin{split}
    \int_0^T \big(\ddt \vn, w \big)_{L^2} \dd t
    &= \int_0^T \big(\Pn(-a\, \CA\, \vn + \CB\, \vn + f), w \big)_{L^2} \dd t \\
    &= \int_0^T \big( -a\, \CA\, \vn + \CB\, \vn + f, w \big)_{L^2}
    \dd t .
  \end{split}
\end{equation}
Setting $n = n_l$ we find in the limit $l \ra \infty$ that
\begin{equation}
  \label{4.9}
  \int_0^T (\dot{v}, w )_{L^2} \dd t = \int_0^T (-a\, \CA\, v + \CB\, v + f, w )_{L^2} \dd t\, .
\end{equation}
Since test functions of this type are dense in $L^2((0,T), L^2 (G))$,
eq.\ (\ref{4.9}) holds for any $w \in L^2((0,T), L^2(G))$. This proves
(\ref{4.2}a) to be an equality in $L^2((0,T), L^2(G))$.

Inserting $w \in C^1([0,T], L^2(G))$ with $w(T) = 0$ in (\ref{4.9}), after integration by parts on the left-hand side 
we find
$$
- \int_0^T \big( v,\ddt w \big)_{L^2} \dd t + (v(0), w(0))_{L^2}\, .
$$ 
Doing the same in (\ref{4.8}) yields in the limit $n_l \ra \infty$ on the left-hand
side,
$$
- \int_0^T \big( v,\ddt w \big)_{L^2} \dd t + (v_0,
w(0))_{L^2}\, .
$$ 
Since $w(0) \in L^2(G)$ is arbitrary we have $v(0) =
v_0$ in $L^2 (G)$. These results prove $v$ to be a weak solution of
problem (\ref{4.2}).

Finally, to prove uniqueness of the weak solution consider $v_1 - v_2
=: v_0$ satisfying
$$
\dot{v}_0 = - a\, \CA\, v_0 + \CB\, v_0\, , \qquad v_0 (0) = 0\, .
$$
Setting in eq.\ (\ref{4.9}) $v:= v_0 \in L^2((0,T), D(\CA))$, $w :=
\CA\, v_0$, and $f=0$ we obtain for a.e. $t \in [0,T]$:
\begin{equation}
  \label{4.11}
  (\dot{v}_0, \CA\, v_0)_{L^2} = - (a\, \CA \, v_0,\CA \, v_0)_{L^2} + (\CB\, v_0, \CA\, v_0)_{L^2}\, .
\end{equation}
The left-hand side of (\ref{4.11}) combined with (\ref{4.A}) takes the form
$$
(\dot{v}_0, \CA\, v_0)_{L^2} = \langle \CA^{1/2} \dot{v}_0, \CA^{1/2}
v_0\rangle = \frac{1}{2}\, \ddt \,\|\CA^{1/2} v_0\|^2_{L^2} =
\frac{1}{2}\, \ddt \,\|v_0\|^2_{1/2}\, ,
$$
whereas estimates analogous to that leading to (\ref{4.5}) show that
the right-hand side of (\ref{4.11}) can be bounded by $\frac{1}{2} C_1
\|v_0\|^2_{1/2}$.  Since $t \mapsto \| v_0 (t) \|_{1/2}$ is absolutely
continuous by lemma 4.2, applying Gronwall to the inequality
$$
\ddt\, \| v_0 \|_{1/2}^2 \leq C_1 \| v_0 \|^2_{1/2} \qquad \mbox{ for
  a.e. } t \in [0,T]
$$
with $\|v_0 (0)\|_{1/2} = 0$ yields the desired result $v_0 \equiv 0$.
\qed
\begin{remark} {\rm
  As to the original (shifted) problem {\rm (\ref{4.1})} theorems 
  3.3 and 4.3 imply that $u_s (x,t) := [\vti (t)] (x)$, where
  $\vti (t)$ is the harmonic extension of $v(t)$, satisfies 
  (\ref{4.1}a,b) for a.e.$\, (x,t) \in \R^d \ti [0,T]$. }
\end{remark}
Higher regularity of the weak solution depends on the smoothness of
the coefficients and the initial-value, and suitable compatibility conditions 
among these data. There holds
\begin{theorem}[Higher regularity]
  Let $T>0$, $k \in \N\setminus\{1\}$, and $a,b,c \in C_1^k(\Go \ti
  [0,T])$.\footnote{In this notation the upper index at ``C" refers to
    the order of spatial derivatives and the lower one (omitted if
    zero) to temporal derivatives; so, $a \in C_1^k (\Go \ti [0,T])$
    means $a$, $\pa_t a$, and $D^\al _x a$, $|\al| \leq k$ are all
    continuous functions on $\Go \ti [0,T]$.}
  Let, furthermore, $v_0 \in D_{(k+1)/2}$, 
$-a(\cdot ,0) \CA v_0 + \CB|_{t = 0} v_0 + f(0) \in D_{(k-1)/2}$, and $f \in C^1([0,T], H^k(G))$. Then the
  weak solution $v$ of problem {\rm (\ref{4.2})} satisfies 
$$
v \in L^2 ((0,T), D_{k/2\, +\, 1})\, ,\quad \dot{v} \in L^2 ((0,T),
D_{k/2})\, ,\quad \ddot{v} \in L^2((0,T), D_{k/2\, -\, 1})\, .
$$
\end{theorem}
{\textsc{Proof:}}
Higher spatial regularity is easily obtained via the operator $\CA$:
applying $\CA^{k/2}$ on (\ref{4.2}) and setting 
$\CA^{k/2} v{=:}w,$
$$
\CA^{k/2} f =: f^{(k)}, \quad \CA^{k/2} \CB\, \CA^{-k/2} =: \CB^{(k)},\quad
(a -\CA^{k/2} a\, \CA^{-k/2}) \CA =: \CC^{(k)},
$$
and $\CA^{k/2} v_0 := w_0$, we obtain
\begin{subequations}
  \label{4.12}
  \begin{align}
    \dot{w}  &= - a\, \CA\, w + \CB^{(k)}\, w + \CC^{(k)}\, w + f^{(k)}\, ,\\
    w(0) &= w_0\, .
  \end{align}
\end{subequations}
$f^{(k)}$ and $w_0$ fulfill the prerequisites of theorem 4.3,
$$
\CB^{(k)} : C([0,T], D_{1/2}) \ra C([0,T], L^2(G))
$$ 
is again a
bounded operator, and $\CC^{(k)}$ is of the same type as
$\CB^{(k)}$. Thus theorem 4.3 applies to (\ref{4.12}) with the result
$$
w \in L^2 ((0,T), D_1)\, ,\qquad \dot{w} \in L^2 ((0,T), L^2(G))\, ,
$$
i.e.
$$
v \in L^2 ((0,T), D_{k/2\, +\, 1})\, ,\qquad \dot{v} \in L^2 ((0,T),
D_{k/2})\, .
$$

To obtain higher temporal regularity we need some more a-priori
estimates for the Galerkin approximations $\vn$. Note that $a,\, b,\,
c$ are at least $\in C_1^2 (\Go \ti [0,T])$, $v_0 \in D_{3/2}$, and $f
\in C^1 ([0,T], D(\CA))$. So, inserting $\wn := \CA \vn $ into
(\ref{4.12}a) and taking the scalar product with $\CA \wn$ we obtain
$$
\frac{1}{2}\, \ddt \,\|\wn\|^2_{1/2} = ( \CA\, \wn, - a\, \CA\, \wn +
\CB^{(2)} \wn + \CC^{(2)} \wn + f^{(2)} )_{L^2} \, .
$$
This is analogous to (\ref{4.4}) and the subsequent estimates leading
to (\ref{4.3}) yield now
\begin{equation}
  \label{4.13}
  \max_{[0,T]} \| \wn \|_{1/2}^2 = \max_{[0,T]} \| \vn \|_{3/2}^2 \leq C_3\, , 
\end{equation}
and, via (\ref{4.B}a),
\begin{equation}
  \label{4.15}
  \max_{[0,T]} \big\| \ddt \vn \big\|_{1/2}^2 \leq C_4\, .
\end{equation}
On the other hand, differentiating (\ref{4.B}a) with respect to $t$,
setting 
$$
\ddt \vn =: \vdn,\quad \ddt f =: \dot{f}\; \mbox{ and}\quad (\pa_t b \cdot
\na \vn + \pa_t c\, \vn ) =: \dot{\CB}\, \vn ,
$$
we obtain
\begin{equation}
  \label{4.C}
  \ddt \vdn  = \Pn \big(- a\, \CA\, \vdn + \CB\, \vdn + \dot{f} - \pa_t a\, \CA \, \vn + \dot{\CB}\, \vn \big)\, ,
\end{equation}
which is of type (\ref{4.B}a). Complementing (\ref{4.C}) by the initial-value $\vdn (0) = \Pn \dot{v}_0$, where 
$\dot{v}_0 := -a(\cdot ,0) \CA v_0 + \CB|_{t = 0} v_0 + f(0) \in D_{1/2}$, we obtain again a bound of type (\ref{4.3}), now on $\vdn$.
Next, we modify for \reff{4.C} the argument which leads from (\ref{4.3})
to (\ref{4.AA}). Taking the scalar product of \reff{4.C} with $\CA\,
\vdn$ and observing that $\dot{\CB}$ is of the same type as $\CB$,
thus using the bounds 
$$
\|\CB\, \vdn \|_{L^2} \leq (C_1 a_0)^{1/2}
\|\vdn \|_{1/2}\, , \qquad \|\dot{\CB}\, \vn \|_{L^2} \leq (C_1 a_0)^{1/2}
\|\vn \|_{1/2} 
$$ 
as well as $\max_{[0,T]} \| \vn \|_{1}^2 \leq C_5$,
\reff{4.3}, and $a\geq a_0$, $|\pa_t a| \leq A$, $\max_{[0,T]}
\|\dot{f}\|_{L^2}^2 \leq C_2 a_0$, we obtain
\begin{equation*}
  \begin{split}
    \frac{1}{2}\, \ddt \,\|\vdn\|^2_{1/2} \leq - &a_0 \| \vdn \|^2_1 + \frac{a_0}{8} \| \vdn \|_1^2 + 2\, C_1 \| \vdn \|_{1/2}^2 + \frac{a_0}{8} \| \vdn \|_1^2 + 2\, C_2 \\
    & + \frac{a_0}{8} \| \vdn \|_1^2 + 2\, \frac{A}{a_0}\, C_5 +
    \frac{a_0}{8} \| \vdn \|_1^2 + 2\, C_1 C\, .
  \end{split}
\end{equation*}
This estimate is analogous to \reff{4.10}, so we have
$$
\int_0^T \|\vdn\|_1^2 \dd t \leq C_6
$$
and, using \reff{4.C} once more,
$$
\int_0^T \big\| \ddt \vdn \big\|_{L^2}^2 \dd t \leq C_7\, .
$$ 
Recalling the reasoning after (\ref{4.7}) there is thus a subsequence
$(n_{l_m}) =: (m)$ of $(n)$ and a function $\ddot{v} \in L^2((0,T),
L^2(G))$ such that $\ddot{v}$ is the weak limit of $\big(\ddt
\vdm\big)$ and the weak time derivative of $\dot{v}$. Finally,
inspecting again \reff{4.C} we find that there is enough regularity of
the data $a$, $b$, $c$, $f$, and $v_0$ left to improve the spatial
regularity of $\ddot{v}$ by the order $k-2$, i.e.\ we have $\ddot{v}
\in L^2((0,T), D_{k/2\, -\, 1})$.  \qed

\vspace{1ex}

In view of lemma 4.2 and Sobolev's embedding theorems, theorem 4.5
implies the existence of smooth solutions. The following corollary
formulates for the original evolution problem (\ref{1.1})
sufficient (and not necessarily sharp) 
conditions in terms of classical derivatives for existence of 
classical solutions. 
In particular, to express the compatibility conditions in classical terms, 
note that $u \in C^2 (\Go)$ and 
$\uti \in  C^1 (\R^d)$ imply $\uti \in H^2_{loc}(\R^d)$. 
\begin{corollary}[Classical solution of the evolution problem]
  Let $G \subset \R^d$, $d \geq 3$ be a bounded domain with $C^{k +
    3/2}$-boundary, $k > 1 + d/2$, $u_0 \in C^{k+1} (\Go)$, and 
$a,b,c \in C_1^k(\Go \ti [0,T])$, $u_\infty \in C^2([0,T])$ for 
any $T> 0$. Let, furthermore, 
$u_0 - u_\infty (0)$, $\Delta^i u_0$, and $\Delta^{i-1} (a_0 \Delta u_0 + b_0 \cdot \na u_0 + c_0 u_0 - \dot{u}_\infty (0))$, $i= 1,\ldots,[(k-1)/2]$, where $a_0 =a(\cdot , 0)$ etc., all $C^1$-match to their harmonic extensions.
  Then problem {\rm (\ref{1.1})}
  has a unique classical solution $u$, i.e.\ $u \in C_1^2 (G \ti \R_+)
  \cap C^2(\Gh \ti \R_+)$ satisfies pointwise eqs.\ {\rm (\ref{1.1})}.
\end{corollary} 
{\textsc{Proof:}}
Fixing some $T>0$, by theorem 3.3, lemma 4.2, and Sobolev's
embedding theorems, from theorem 4.5 we find
\begin{equation*}
  \begin{split}
    v &\in C ([0,T], H^{k+1} (G)) \subset C([0,T], C^2 (\Go))\, , \\
    \dot{v} &\in C ([0,T], H^{k-1} (G)) \subset C([0,T], C (\Go))\, .
  \end{split}
\end{equation*}
Thus, setting $u_s (x,t) := [\vti (t)] (x)$ with $\vti (t)$ being the
harmonic extension of $v(t) \in D_{1/2}$, we have $u_s \in C_1^2 (\Go
\ti [0,T])$ satisfying (\ref{4.1}a) and (\ref{4.1}e). Since $\vti (t)
\in \CH_0$ is harmonic in $\Gh$ conditions (\ref{4.1}b) and
(\ref{4.1}d) hold for $u_s$ as well. To prove (\ref{4.1}c) note that
this condition holds for $\vti_n$. Recalling the maximum principle for
harmonic functions this implies
$$
\max_{x \in \R^d} |D_x^\al \vti_n (x)| = \max_{x \in \Go} |D_x^\al
\vti_n (x)|
$$ 
for any multiindex $\al$ with $|\al| \leq 1$.  Setting $S_{mn} (x,t)
:= \sum_{\nu = m}^n c_\nu (t)\, \vti_\nu (x)$ we have then with
Sobolev and relation (\ref{4})
\begin{equation}
  \label{4.E}
  \begin{split}
    \max_{t \in [0,T]} \sum _{|\al| \leq 1} \max_{x \in \R^d} |D_x^\al S_{mn}(x,t) | =&  \max_{t \in [0,T]} \sum _{|\al| \leq 1} \max_{x \in \Go} |D_x^\al S_{mn}(x,t) | \\
    \leq&C \max_{t \in [0,T]} \| S_{mn}(\cdot ,t) \|_{H^k (G)} \leq
    \widetilde{C} \| S_{mn} \|_{C([0,T], D_{k/2})}\, .
  \end{split}
\end{equation}  
So, convergence of $v (t) = \sum_{n=1}^\infty c_n(t)\, v_n$ in
$C([0,T],D_{k/2})$ implies convergence of $\vti (t) =
\sum_{n=1}^\infty c_n(t)\, \vti_n$ in $C([0,T], C^1 (\R^d))$, i.e.\
$u_s \in C^1(\R^d \ti [0,T])$.

Similarly, fixing any $K\Subset \Gh$ and using the interior derivative
estimate (cf.\ \cite[p23]{gt77})
$$
\max_{x \in K} |D_x^\al \vti_n (x)| \leq \widehat{C} \max_{x \in
  \overline{\Gh}} |\vti_n (x)| \leq \check{C} \max_{x \in \Go} |\vti_n
(x)|
$$ 
with $|\al | = 2$ we find $u_s \in C^2(K \ti [0,T])$ and hence $u_s
\in C^2(\Gh \ti [0,T])$.

Observing, finally, that $T>0$ is arbitrary
we have in conclusion that $u_s$ is a classical solution of problem (\ref{4.1}) and hence
$u:= u_s + u_\infty$ is a classical solution of problem (\ref{1.1}).  
\qed
\begin{remark}{\rm
In $d= 3$ we may choose $k=3$ and the compatibility conditions amount to 
\begin{equation}
\label{a}
 \widetilde{u_0 - u_\infty (0)}\, , \, \widetilde{\Delta u_0}\, \in C^1(\R^3)
\end{equation}
and 
\begin{equation}
\label{b}
(a_0 \Delta u_0 + b_0 \cdot \na u_0 + c_0 u_0 - \dot{u}_\infty(0))\widetilde{\hspace{5mm}} \in C^1(\R^3)\, .
\end{equation}
So, in the case $u_\infty = 0$ admissible initial values $u_0$ are 
for instance $C^4(\Go)$-functions with 
$\pa^i_n u_0\big|_{\pa G} = 0$, $i=0, \ldots, 3$, where $\pa_n$ denotes the normal derivative at $\pa G$. In the case $u_0 = u_\infty = const >0$, which was interesting in applications \cite{kai07}, condition (\ref{b}) requires the coefficient $c_0$ to have a $C^1$-smooth harmonic extension. 
}
\end{remark}
%
%

\vspace{5ex}

\noindent {\bf{\Large Appendices}}
\setcounter{section}{0} \renewcommand{\thesection}{\Alph{section}}
\section{The non-radial-flow problem}
In the framework of magnetohydrodynamics the kinematic dynamo problem
reads \cite{backus58}:
\begin{subequations}
  \label{AA.1}
  \begin{alignat}{2}
    \pa_t B + \na\ti (\eta \na \ti B) &= \na \ti (v \ti B)\, , \quad \na \cdot B = 0 &&\qquad \mbox{ in } G \ti \R_+,\\
    \na \ti B &= 0\, , \quad \na \cdot B = 0 &&\qquad \mbox{ in } \Gh \ti \R_+,\\
    B &\mbox{ continuous } &&\qquad \mbox{ in } \R^3 \ti \R_+, \\
    B(x,t) &= O(|x|^{-3}) &&\qquad \mbox{ for } |x| \ra \infty,\, t \in \R_+ , \\
    B(\cdot,0) &= B_0 &&\qquad \mbox{ on } G \ti \{t = 0\}.
  \end{alignat}
\end{subequations}
Here, the induction equation (\ref{AA.1}a) describes the generation of
the magnetic field $B$ by the motion (with prescribed flow field $v$)
of a conducting fluid (with conductivity $\eta > 0$) in a bounded
region $G \subset \R^3$. Outside the fluid region there are no further
sources of magnetic field. Thus, $B$ continues in $\Gh = \R^3
\setminus \Go$ as a vacuum field and vanishes at spatial infinity.

If $G$ is a ball $B_R$ (or a spherical shell) the so-called
poloidal-toroidal decomposition of solenoidal fields is especially
useful \cite{backus58,schmitt95}: 
$$
B = B_P + B_T = - \na \ti \Lambda\, S - \Lambda\, T\, ,\qquad \langle
S\rangle = \langle T \rangle = 0.
$$
$\Lambda$ denotes here the non-radial derivative operator $\Lambda :=
x \ti \na$, $\Lambda \cdot \Lambda =:\CL$ is the
Laplace-Beltrami-operator on the unit sphere $S_1$, and $\langle
\,\cdot\, \rangle$ denotes the spherical mean. The poloidal and
toroidal scalars $S$ and $T$, resp., are uniquely determined (e.g.\ in
$L^2 (B_R)$) by $B$:
$$
x \cdot B = - \CL\, S\, , \qquad x\cdot \na \ti B = - \CL\, T.
$$
In the following we refer to $P:= x\cdot B$ instead of $S$ as the poloidal
scalar.

In the case of a non-radial flow field, i.e.\ $v\cdot x \equiv 0$, and
spherically symmetric conductivity, problem (\ref{AA.1}) implies the
scalar sub-problem (\ref{1.3}) for $P$: (\ref{1.3}a) is just the
radial component of the first part of (\ref{AA.1}a) and (\ref{1.3}b) is
obtained by applying $\Lambda$ on the first part of
(\ref{AA.1}b). Condition (\ref{1.3}c) is in fact enough to ensure a
continuous magnetic field $B$, i.e.\ continuous second-order
derivatives of $S = - \CL^{-1} P$, since $B$ involves not more than
one radial derivative of $S$. The equivalence of (\ref{AA.1}d) with
(\ref{1.3}d) is clear for $B\cdot (x/|x|)$ and follows for the
non-radial components with the divergence-constraint.
\begin{remark}
{\rm In the mathematical treatment of problem (\ref{1.1}) it turns out to be useful to consider functions $v$ satisfying the integral condition $\int_{\R^d} |\na v|^2 \dd x < \infty$. In the context of problem (\ref{1.3}) this condition can be interpreted as one guaranteeing finite total magnetic energy. In fact, the total energy of the poloidal magnetic field reads 
$$
E[B_P] = \frac{1}{2} \int_{\R^3} |\na \ti \Lambda\, S|^2 \dd x = \frac{1}{2} \int_{\R^3} |\na \CL^{1/2} S|^2 \dd x,
$$
and with the variational estimate 
$$
\inf_{f \neq 0, \langle f \rangle = 0} \frac{\|\CL\, f\|_{L^2 (S_1)}}{\|f\|_{L^2(S_1)}} = 2 ,
$$
one obtains the bound on $E[B_P]$:
$$
E[B_P] \leq \frac{1}{4} \int_{\R^3} |\na \CL\, S|^2 \dd x = \frac{1}{4} \int_{\R^3} |\na P |^2 \dd x.
$$
}
\end{remark}
\section{The axisymmetric problem}
The central assumption is here an axisymmetric magnetic field with
representation
\begin{equation}
  \label{B.1}
  B = \na P \ti \na \phi + A \na \phi = - \frac{1}{\rho}\, \pa_z P\, \er_\rho + \frac{1}{\rho}\, A\, \er_\phi + \frac{1}{\rho}\, \pa_\rho P\, \er_z
\end{equation}
by two scalar quantities, the poloidal one $P$ and the toroidal or
azimuthal one $A$, depending (besides on $t$) on $\rho$ and $z$ with
$(\rho,\phi,z)$ being cylindrical coordinates in $\R^3 \setminus\{\rho
= 0\}$. Inserting (\ref{B.1}) into the dynamo equation (\ref{A.1}) the
following sub-problem for the poloidal scalar $P$ arises \cite{backus57,
lms82, ij84}:
\begin{subequations}
  \label{B.2}
  \begin{alignat}{2}
    \pa_t P - \eta \Delta_* P &= - v\cdot \nab P &&\qquad \mbox{ in } G_2 \ti \R_+,\\
    \Delta_* P &= 0 &&\qquad \mbox{ in } \Gh_2 \ti \R_+,\\
    P \mbox{ and } \na P &\mbox{ continuous } &&\qquad \mbox{ in } H \ti \R_+, \\
    P(\rho,z,t) &\ra 0 &&\qquad \mbox{ for } \rho \ra 0,\, (z,t) \in \R \ti \R_+ , \\
    |\na P(\rho,z,t)| &= O(\rho) &&\qquad \mbox{ for } \rho \ra  0 ,\, (z,t) \in \R \ti \R_+, \\
    \Big|\frac{1}{\rho} \na P(\rho,z,t)\Big| &= O\big((\rho^2 + z^2)^{-3/2}\big) &&\qquad \mbox{ for } \rho^2 + z^2 \ra \infty, \, t \in \R_+, \\
    P(\cdot,\cdot,0) &= P_0\, ,\quad P_0 \mbox{ satisfying (B.2d,e)}
    &&\qquad \mbox{ on } G_2 \ti \{t = 0\}.
  \end{alignat}
\end{subequations}
$\Delta_*$ is the elliptic operator $\Delta_* := \pa_\rho^2 -
\frac{1}{\rho} \pa_\rho + \pa_z^2$ on the half-plane $H:= \R_+ \ti
\R$.  $G_2 \subset H$ is the ``cross-section" of some bounded region
$G_3 \subset \R^3$; more precisely, $\ol{G_3} \setminus\pa\ol{G_3}$
with $G_3 := G_2 \ti \{0\leq \phi < 2\pi\}$ is a bounded domain in
$\R^3$ with smooth boundary. Note that the axisymmetric flow field
need not be solenoidal, the azimuthal component, however, w.l.o.g.\
can assumed to be zero. Condition (\ref{B.2}e) ensures a finite
magnetic field on the symmetry axis $\{\rho = 0\}$. It implies the
limit $\lim_{\rho \ra 0} P(\rho,\cdot,t) = P_s (t)$, where $P_s$
depends only on $t$. As $P_s$ does not affect the magnetic field it is
set to zero for simplicity (condition (\ref{B.2}d)). Note that in 
\cite{lms82, ij84} conditions
(\ref{B.2}d,e) are replaced by
\begin{equation}
  \label{B.3}
  P(\rho,z,t) = O(\rho^2)\qquad \mbox{ for } \rho \ra 0,\, (z,t) \in \R \ti \R_+ . 
\end{equation}
The cautious Backus \cite{backus57} requires (\ref{B.2}e) {\em and}
(\ref{B.3}). In fact, conditions (\ref{B.2}d,e) imply (\ref{B.3}), but
not vice versa. In the view of the original problem (\ref{A.1}),
condition (\ref{B.2}e) seems to be the natural one. Similarly, in
these references the ``natural" condition (\ref{B.2}f) is replaced by
\begin{equation}
  \label{B.4}
  P(\rho,z,t) = O\big((\rho^2 + z^2)^{-1/2}\big)\qquad \mbox{ for } \rho^2 + z^2 \ra \infty,\, t \in \R_+ . 
\end{equation}
These conditions are in fact equivalent for solutions of (\ref{B.2}b)
as becomes clear in the subsequent formulation of problem (\ref{B.2}).

There is an elegant way to eliminate the ``coordinate--singularity" at
$\rho = 0$ in problem (\ref{B.2}), namely by embedding (\ref{B.2}) in
$\R^5$. $P$ is then considered as an axisymmetric function in $\R^5$
with symmetry axis in $x_5$-direction. Identifying $\rho^2$ with
$\sum_{i=1}^4 x_i^2$ and $z$ with $x_5$, $x\in \R^5$, and introducing
$Q(x,t) := \Qti (\rho,z,t) := P(\rho,z,t)/\rho^2$ the crucial
observation is \cite{chand56} 
$$
\Delta_* P = (\pa_\rho^2 - \frac{1}{\rho}\, \pa_\rho + \pa_z^2) P =
\rho^2 (\pa_\rho^2 + \frac{3}{\rho}\, \pa_\rho + \pa_z^2) \Qti =
\rho^2 \Delta_5 Q
$$
with $\Delta_5$ being the Laplacian in $\R^5$. With the further
definitions
\begin{alignat*}{2}
  b_i (x,t) &:=- v_\rho (\rho,z,t)\, \frac{x_i}{\rho}\, ,\; i=1,\ldots ,4, &\qquad b_5(x,t) &:= - v_z (\rho,z,t). \\
  c (x,t) &:=- 2\, v_\rho (\rho, z,t) /\rho\, , &\qquad a(x,t) &:=
  \eta (\rho,z,t)
\end{alignat*}
problem (\ref{B.2}) takes in $\R^5$ the form
\begin{subequations}
  \label{B.5}
  \begin{alignat}{2}
    \pa_t Q - a \Delta_5 Q &= b \cdot \nab Q + c\, Q &&\qquad \mbox{ in } G_5 \ti \R_+,\\
    \Delta_5 Q &= 0 &&\qquad \mbox{ in } \Gh_5 \ti \R_+,\\
    Q \mbox{ and } \na Q &\mbox{ continuous } &&\qquad \mbox{ in } \R^5 \ti \R_+, \\
    Q(x,t) &= O(|x|^{-3}) &&\qquad \mbox{ for } |x| \ra \infty, \, t \in \R_+, \\
    Q(\cdot,0) &= Q_0\, ,\quad Q_0 \mbox{ axisym. } &&\qquad \mbox{ on
    } G_5 \ti \{t = 0\}.
  \end{alignat}
\end{subequations}
$G_5$ is now an axisymmetric bounded region in $\R^5$. An axisymmetric
initial field $Q_0$ implies axisymmetry of $Q(\cdot, t)$ for all
$t>0$. A condition on the symmetry axis is no longer necessary;
conditions (\ref{B.2}d,e) (as well as (\ref{B.3})) are automatically
satisfied by $P:= \rho^2 \Qti$. As to the behaviour for large $x$, $Q$
is in $\Gh$ an exterior harmonic function with representation
(\ref{C.3}); thus, (\ref{B.5}d) implies $|\na Q (x,\cdot)| =
O(|x|^{-4})$ for $|x| \ra \infty$ and hence (\ref{B.2}f) as well as
(\ref{B.4}). On the other side, in the view of (\ref{C.6}) each of
the conditions (\ref{B.2}f) and (\ref{B.4}) implies (\ref{B.5}d).
\begin{remark} {\rm Stredulinsky et al.\ \cite{sml86} doubt the correctness
    of the boundary condition (\ref{B.4}) in \cite{ij84} 
    and cite their own results (theorem 2) about solutions with
    nonvanishing asymptotic value $P_\infty(t)$ at spatial
    infinity. They mention the possibility of $\lim_{t \ra \infty}
    P_\infty (t) \neq 0$, even when $\lim_{t \ra 0} P_\infty (t) =
    0$. In fact, condition (\ref{B.4}) is correct as demonstrated
    above, which means $P_\infty \equiv 0$ in the axisymmetric
    problem. The discrepancy arises because in \cite{sml86} the authors are
    especially interested in the two-dimensional case where their
    problem (1) makes physical sense (``dynamo problem with plane
    symmetry").  In $d=2$, in fact, $P_\infty \neq 0$ cannot be
    avoided in general. In $d>2$, however, problem (1) is
    underdetermined and the condition $P_\infty \equiv 0$ can be
    added. Observe in this context that in \cite{sml86} the authors do not
    claim uniqueness for solutions of problem (1), uniqueness is claimed only for weak
    solutions which are ``minimizers" (this is more than ``harmonic")
    in $\Gh$ (theorem 1).  }
\end{remark}
\section{Exterior harmonic functions}
An exterior harmonic function $u$ is called ``harmonic at infinity"
iff $u(x) \ra 0$ for $|x| \ra \infty$. In $d \geq
3$ dimensions these functions have the series representation
\begin{equation}
  \label{C.3}
  u(x) = \sum_{n=0}^\infty |x|^{2-n-d}\, Y_n (x/|x|)\, , \qquad Y_n \in H_n\, ,
\end{equation}
absolutely and uniformly converging in the exterior of any ball $B_r
\subset \R^d$ with $r>1$ (cf.~\cite[p115]{folland95}. $H_n$ is the
space of all harmonic homogeneous polynomials of degree $n$ in $\R^d$
restricted to the unit sphere $S^{d-1}$ with dimension $D_n := \dim
H_n =(2n + d -2) (n+d-3)![n!(d-2)!]^{-1}$ (cf.~\cite[p98f]{folland95}. 
In particular, $\dim H_0 = 1$ and $Y_0 = const$; any other
$Y_n$ has vanishing spherical mean, $\langle Y_n\rangle = 0$, $n \in
\N$. The total of spaces $H_n$ spans $L^2(S^{d-1})$: $L^2(S^{d-1}) =
\bigoplus_0^\infty H_n$. So, choosing orthonormal bases $\{Y_{nm}\,
|\, 1\leq m\leq D_n\}$ in $H_n$, any $f \in L^2(S^{d-1})$ allows a
unique representation
\begin{equation}
  \label{C.4}
  f = \sum_{n=0}^\infty \sum_{m=1}^{D_n} c_{nm}\, Y_{nm}
\end{equation}
with coefficients $c_{nm} := (f,Y_{nm})_{L^2} \in \R$. Obviously,
(\ref{C.4}) is the higher-dimensional analogue of the well-known
representation
\begin{equation}
  \label{C.5}
  f = \sum_{n=0}^\infty \sum_{m=-n}^{n} c_{nm}\, Y_{nm}
\end{equation}
by spherical harmonics $\{ Y_{nm}\, |\, n \in \N_0,\, |m| \leq n\}$ in
$d=3$. Note that the $Y_{nm}$ are in our setting real quantities; by
taking suitable linear combinations this is true for spherical
harmonics as well.

Exterior harmonic functions $u$ satisfying the condition $\int_{\Bh_R}
|\na u|^2 \dd x < \infty$ for some $R>0$ are harmonic at infinity up
to a constant $c_0$ (cf.~\cite[p41]{simsohr96}). With
(\ref{C.3}) this implies the representation
\begin{equation}
  \label{C.6}
  u(x) = c_0 + \sum_{n=0}^\infty |x|^{2-n-d}\, Y_n (x/|x|)\, , \qquad c_0 \in \R\, ,\quad Y_n \in H_n\, .
\end{equation}
(\ref{C.6}) holds also in the case of exterior harmonic functions with
asymptotic conditions $u(x) = O(1)$ for $|x| \ra \infty$ 
(cf.~\cite[p64]{axler92}) or $|\na u| = O(|x|^{1-d})$ for $|x| \ra
\infty$. The latter statement follows from the former and the
estimate
$$
|u(x)| \leq |u(R\, x/|x|)| + \int_R^{|x|} |\na u(r\, x/|x|)| \dd r\, ,
\qquad |x| \geq R\geq 0\, .
$$
Let us, finally, consider the exterior boundary-value problem
\begin{subequations}
  \label{C.1}
  \begin{alignat}{2}
    \Delta u &= 0 &&\qquad \mbox{ in } \Gh ,\\
    u &= \phi &&\qquad \mbox{ on } \pa \Gh ,\\
    u(x) &\ra c &&\qquad \mbox{ for } |x| \ra \infty .
  \end{alignat}
\end{subequations}
Here, $\Gh \subset \R^d$, $d\geq 3$ is an exterior region, i.e.\ $\Gh
= \R^d \setminus \Go$ for some bounded domain $G\subset \R^d$, with
$C^1$-boundary $\pa \Gh$. For $\phi \in C(\pa \Gh)$ and $c\in \R$ we
call $u \in C(\Gh \cup \pa \Gh) \cap C^2 (\Gh)$ satisfying (\ref{C.1})
a classical solution. A weak formulation is based on the spaces $\Hh
:= \{ v\in H^1_{loc} (\Gh) \, |\, \|v\|_{\Hh} < \infty\}$ and $\Hh_0
:= \mbox{clos} \{ C_0^\infty (\Gh)\, ,\, \| \cdot \|_{\Hh} \}$ with
$\|v\|^2_{\Hh} := \int_{\Gh} |\na v|^2 \dd x$. Describing the boundary
and asymptotic conditions by a function $w \in \Hh$, a weak version of
(\ref{C.1}) reads:
\begin{subequations}
  \label{C.2}
  \begin{align}
    \int_{\Gh} \na u\cdot \na v \dd x &= 0 \qquad \mbox{for any }\, v \in \Hh_0 , \\
    u - w &\in \Hh_0\, .
  \end{align}
\end{subequations}
For problems (\ref{C.1}), (\ref{C.2}) holds:
\begin{theorem}[Solution of the exterior Dirichlet problem] The
  exterior boun\-da\-ry-value problem {\rm (\ref{C.2})} with given $w \in
  \Hh$ has a unique solution $u \in \Hh$. Moreover, $u \in C^\infty
  (\Gh)$ and $\Delta u = 0$ in $\Gh$. If $w|_{\pa \Gh} = \phi \in
  C(\pa \Gh)$ and $w \ra c$ for $|x| \ra \infty$, then $u$ is a
  classical solution of {\rm (\ref{C.1})}. Furthermore, $u$ is the
  unique minimizer of the functional $\|\cdot\|_{\Hh}$ on the set $\{v
  + w\, |\, v \in \Hh_0 \}$.
\end{theorem}
For a proof we refer to \cite[p543]{dl90}. We note
only that $\|\cdot\|_{\Hh}$-convergence already implies
$\|\cdot\|_{L^2_{loc} (\Gh)}$-convergence, thus $\Hh_0 \subset
\Hh$. In fact, the Gagliardo-Nirenberg-Sobolev-inequality 
(cf.~\cite[p263]{evans}) implies the estimate
$$
\| v\|_{L^p (\Gh)} \leq C \| v \|_{\Hh}\, , \qquad p = \frac{2\, d}{d
  - 2}
$$
for any $v \in \Hh_0$ and a constant $C$ depending only on $d$. So,
fixing some $K \Subset \Gh$ we obtain:
$$
\| v\|_{L^2 (K)} \leq C(K) \| v\|_{L^p (K)} \leq C(K) \| v\|_{L^p
  (\Gh)} \leq C(K) C \| v \|_{\Hh}\, .
$$
In particular, our $\Hh_0$ coincides with the corresponding space
$B_0^1 (\Gh)$ in \cite{dl90}. 
\section{Poloidal free decay modes}
The poloidal free decay modes are a countable set of explicit
solutions of the eigenvalue problem (\ref{3.1}) if $G$ is a ball $B_R$
in $\R^3$. In terms of spherical Bessel functions $j_n$ and spherical
harmonics (cf.\ appendic C) they take the form
$$
\pti_{lnm} (x) := \sqrt{\frac{2}{R^3}} \left\{
  \begin{array}{ll}
    \ds\frac{j_n(i_l^{n-1} |x|/R)}{j_n (i_l^{n-1})} \, Y_{nm} (x/|x|) & \quad\mbox{ in } B_R \\[1em]
    \ds (|x|/R)^{-n-1} Y_{nm} (x/|x|) & \quad\mbox{ in } \Bh_R
  \end{array} \right. \, ,\quad l \in \N, n \in \N_0, |m| \leq n 
$$
with eigenvalues $\lambda_{lnm} := \lambda_{ln} := (i_l^{n-1}/R)^2$;
$i_l^{n}$ is the $l$-th positive zero of $j_n$. For their restrictions
$p_{lnm} := \pti_{lnm}|_{B_R}$ holds
\begin{theorem}
  The set of functions $\{p_{lnm} : B_R \ra \R\, |\, l \in \N, n \in
  \N_0, |m| \leq n \}$ constitutes a complete orthonormal system in
  $L^2 (B_R)$.
\end{theorem} {\textsc{Proof:}} The orthonormality of the $p_{lnm}$
can be checked by explicit calculation using the orthonormality of the
$Y_{nm}$ and the corresponding relation for the $j_n$ 
(cf.~\cite[p485, eq.~11.4.5]{as64}. To prove the completeness we
show any solution of problem (\ref{3.1}) being a linear combination of
the $\pti_{lnm}$. Theorem 3.1 yields then the completeness of the
$p_{lnm}$. So, let $u$ be a solution of (\ref{3.1}) with eigenvalue
$\lambda >0$. The representation (\ref{C.5}) for $L^2$-functions on
the unit sphere implies the representation
\begin{equation}
  \label{D.1}
  u(x) = \sum_{n=0}^\infty \sum_{m=-n}^{n} u_{nm}(|x|)\, Y_{nm}(x/|x|)
\end{equation}
for $u$ with coefficients $u_{nm}$ depending only on $|x|$. For
$|x|>R$, (\ref{C.3}) yields
$$
u_{nm} (|x|) = c_{nm} |x|^{-n-1}
$$
with $c_{nm} \in \R$, whereas inserting (\ref{D.1}) into (\ref{3.1}a)
yields the differential equation for $u_{nm}$ on $(0,R)$:
\begin{equation}
  \label{D.2}
  -\,\frac{1}{r^2} \ddr \Big(r^2 \ddr\, u_{nm}\Big) + \frac{1}{r^2}\, n(n+1) u_{nm} = \lambda\, u_{nm}
\end{equation}
with $r:= |x|$. After rescaling $s:=\sqrt{\lambda}\, r$, (\ref{D.2})
takes the spherical form of Bessel's differential equation for
$v_{nm}(s):= u_{nm}(s/\sqrt{\lambda})$:
$$
\Big( s^2 \frac{\rm d^2}{{\rm d}s^2} + 2 s \dds + s^2 - n(n+1)\Big)
v_{nm} = 0
$$
with (in $s=0$) regular solutions $j_n$, $n \in \N_0$. Matching inner
and outer solutions according to (\ref{3.1}c) yields the condition
\begin{equation}
  \label{D.3}
  \Big( s \dds j_n (s) + (n+1) j_n (s) \Big) \Big|_{s = \sqrt{\lambda} R} = (s j_{n-1} (s)) \big|_{s = \sqrt{\lambda} R} = 0
\end{equation}
fixing the eigenvalue at
\begin{equation}
  \label{D.4}
  \lambda = \lambda_{ln} = (i_l^{n-1} /R)^2 .
\end{equation}
Note that (\ref{D.3}) holds also in the case $n=0$ with $j_{-1} = \cos
s/s$.

According to theorem 3.1, the eigenvalue $\lambda$ is of finite
multiplicity. Thus, only finitely many pairs $(l,n)$ satisfy
(\ref{D.4}), and (\ref{D.1}) is in fact a (finite) linear combination
of the $\pti_{lnm}$.  \qed
\begin{remark} {\rm The condition of vanishing spherical mean
    eliminates all $n=0$ modes; thus, $\{p_{lnm} \, |\, l \in \N, n
    \in \N, |m| \leq n \}$ is a complete orthonormal system in $\{v
    \in L^2 (B_R)\, |\, \langle v\rangle = 0 \}$.  }
\end{remark}
%
%
%
%
\section*{Acknowledgements}
Discussions with and comments from M. Seehafer, C. G. Simader and
W. von Wahl are gratefully acknowledged.
%
%
\renewcommand{\refname}{References}

\end{document}